\documentclass[
twocolumn,
nofootinbib,
nobibnotes,
amsmath,amssymb,
prb,
]{revtex4-2}

\usepackage{graphicx}
\usepackage{dcolumn}
\usepackage{bm}
\usepackage{hyperref}
\usepackage[mathlines]{lineno}

\usepackage{tikz}
\usetikzlibrary{circuits.ee.IEC}
\usetikzlibrary{calc}
\usetikzlibrary{arrows.meta, positioning, decorations.pathreplacing, decorations.pathmorphing}

\numberwithin{equation}{section}

\newcommand{\ts}{\textstyle}

\begin{document}

\preprint{}

\title{\textbf{
  Scattering and Schmid-Bulgadaev Transition in Qubits Based on Josephson Junction Chains} 
}%

\author{Sviatoslav D. Krasin}
 \affiliation{Moscow Institute of Physics and Technology (MIPT), 141702 Dolgoprudny, Russia}
 \email{krasin.sd@phystech.edu}

\author{Alexey Radkevich}%
 \affiliation{I.E. Tamm Department of Theoretical Physics, P.N. Lebedev Physical Institute, 119991 Moscow, Russia}
 \email{radkevichaa@lebedev.ru}
 
\author{Emil Rizvanov}
 \affiliation{Department of Experimental Physics, Comenius University, SK-84248 Bratislava, Slovakia}
 \email{emil.rizvanov@fmph.uniba.sk}
 
\author{Andrew G. Semenov}%
 \affiliation{I.E. Tamm Department of Theoretical Physics, P.N. Lebedev Physical Institute, 119991 Moscow, Russia}
 \email{semenovag@lebedev.ru}

\date{\today}

\begin{abstract}
  A Josephson junction subject to an ohmic dissipative environment undergoes a zero-temperature quantum phase transition called Schmid-Bulgadaev transition form an insulating state to a superconducting state upon increasing the impedance of the environment above the resistance quantum. Superconducting qubits based on Josephson junction chains represent a system where it is possible to dynamically explore both sides of this transition by probing the scattering of individual plasma modes of the junction chain on a SQUID at the boundary which was achieved in recent experiments. Here, we provide a theoretical analysis of the scattering in the system paying special attention to the non-perturbative superconducting regime. We find that in this regime, a specific diagrammatic resummation must be performed in order to correctly account for fluctuation renormalization of the SQUID's potential. We find analytic expressions for the modes' frequency shifts and quality factors which together provide a full description of scattering. We argue that the non-perturbative nature of the superconducting phase manifests itself through a power-law dependence of the frequency shifts on the SQUID flux and through non-monotonous dependence of the quality factors on the mode frequency. Our predictions may be readily verified in modern low-temperature experiments.
\end{abstract}

\maketitle

\tableofcontents

\section{Introduction}
The properties of superconducting Josephson junctions (JJ) are known to depend significantly on the dissipative environment of the system. One of the most prominent signatures of this influence is the Schmid--Bulgadaev transition --  a zero-temperature, dissipation-induced quantum phase transition in a resistively shunted JJ. It separates a non-superconducting delocalized phase, where phase tunneling is relevant and Cooper-pair transport is suppressed, from a localized superconducting phase, where the phase difference across the Josephson junction is pinned. The transition occurs when the impedance of the ohmic dissipative environment crosses the superconducting resistance quantum \cite{Schmid1983,Bulgadaev1984,SchonZaikin1990},
\begin{equation}
    R_Q = \frac{h}{(2e)^2} \approx 6.45 \text{ kOhm}.
\end{equation}

The first experimental route to detecting the Schmid--Bulgadaev transition was based on transport measurements in a resistively shunted single Josephson junction. In this approach, one measures the differential resistance
\(R=\mathrm d V/\mathrm d I\) \cite{Penttila2001} or differential conductance \(G = \mathrm d I/\mathrm d V\) \cite{Subero2026}. The transition was identified by a distinct change in the differential conductance or resistance at zero bias, which shifted from a peak in the superconducting state to a dip in the insulating state as the environmental dissipation was increased.
\cite{Penttila2001,HakonenSonin2021,Subero2026}.

A second class of experiments realizes the dissipative environment not as a conventional resistor, 
but as a long high-impedance transmission line formed from thousands of JJ (or JJ chains) 
\cite{Kuzmin2019NP,Kuzmin2019NPJ,Kuzmin2025,Kuzmin_scat} hosting sound-like soft plasma modes; 
or a long ultrathin superconducting nanowire \cite{arutyunov} where the ohmic dissipative environment 
is formed by the sound-like Mooij-Sch\"on modes \cite{mooij}. 
In the latter setting, the dissipative environment was shown to affect the transport properties of 
JJs \cite{arutyunov, ALRSZ} through a fluctuation-induced modification of the quasiparticle density of states 
\cite{RSZ17}. The equilibrium current passing through the system was predicted to be affected 
by the environment even more dramatically, completely vanishing in the insulating phase \cite{RSZ19}.

Recent experiments on JJ-based qubits opened a pathway to probe the system spectroscopically rather than through dc transport: one studies how the junction scatters the modes of the electromagnetic environment. On the superconducting side, the junction behaves as an inductive,  short-circuit termination, whereas on the insulating side it behaves as a capacitive,  open-circuit termination. The transition can therefore be detected through a sign change in the mode-frequency shifts, as well as through inelastic photon scattering \cite{Kuzmin_scat,Kuzmin2025}.

Theoretically, the Schmid--Bulgadaev transition in such systems is often investigated by modeling a transmon qubit coupled to an infinite transmission line. Within this framework, proposed signatures of the transition include characteristic features in the fluorescence spectrum and direct photon scattering from phase slips \cite{Houzet2020,Burshtein2020,Burshtein2024}. Evidence for the transition has also been obtained numerically from calculations of the energy spectrum \cite{Paris_diaganalization}.

In this work, we employ the Keldysh formalism \cite{Kamenev2023} to describe dynamic effects 
in the system under consideration. We aim to capture the physics of the system on both 
sides of the Schmid-Bulgadaev transition, including the superconductive phase where simple perturbation theory 
is inapplicable. We also highlight the dependence on the size of the JJ chain in the superconducting 
phase where ''short'' and ''long'' chains exhibit different properties, similarly to \cite{RSZ19}. 
This length dependence may be important for future applications of such devices \cite{Pechenezhskiy2020}.

\section{Model and Formalism}

\subsection{System under Consideration}
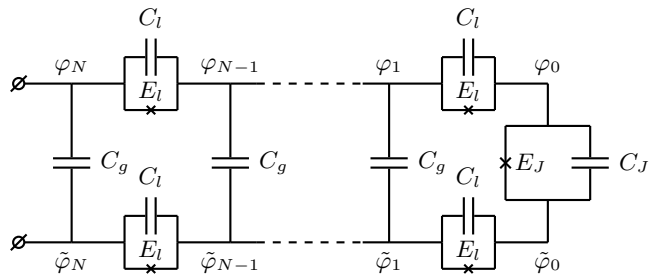
\begin{figure}
    \centering
    \begin{tikzpicture}[circuit ee IEC, thick, scale=0.7]
        \newcommand{\JJ}[2]{
            \draw ($#1!0!#2$) -- ($#1!1!#2$);
            \coordinate (center) at ($#1!0.5!#2$);
            \draw ($(center) - (0.08,0.08)$) -- ($(center) + (0.08,0.08)$) node[midway, above] {$E_l$};
            \draw ($(center) - (0.08,-0.08)$) -- ($(center) + (0.08,-0.08)$);
        }
        \newcommand{\paralnode}[1]{
            \begin{scope}[shift={($#1$)}]
                \draw[thick] (0,0) -- (1,0);
                \draw[thick] (1,0) -- (1,0.5);
                \draw[thick] (1,0) -- (1,-0.5);
                \draw (1,0.5) to [capacitor={info=$C_l$}] (2,0.5);
                \JJ{(1,-0.5)}{(2,-0.5)};
                \draw[thick] (2,0.5) -- (2,0);
                \draw[thick] (2,-0.5) -- (2,0);
                \draw[thick] (2,0) -- (3,0);
            \end{scope}
        }
        
        \newcommand{\crossedterm}[1]{
            \draw[thick, fill=white] (#1) circle (3pt);
            \draw[thick] ($(#1) + (225:6pt)$) -- ($(#1) + (45:6pt)$);
        }

        \draw[thick] (0,0) -- (1,0);
        \crossedterm{0,0}
        
        \draw[thick] (0,-3) -- (1,-3);
        \crossedterm{0,-3}

        \draw (1,0) node[above] {$\varphi_N$};
        \draw (1,-3) node[below] {$\tilde{\varphi}_N$};
        \draw (1,0) to [capacitor={info=$C_g$}] (1,-3);
        \paralnode{(1,0)};
        \paralnode{(1,-3)};
        \draw (4,0) node[above] {$\varphi_{N-1}$};
        \draw (4,-3) node[below] {$\tilde{\varphi}_{N-1}$};       
        \draw (4,0) to [capacitor={info=$C_g$}] (4,-3);
        \draw[thick] (4,0) -- (4.5,0);
        \draw[dashed] (4.5,0) -- (6.5,0);
        \draw[thick] (6.5,0) -- (7,0);
        \draw[thick] (4,-3) -- (4.5,-3);
        \draw[dashed] (4.5,-3) -- (6.5,-3);
        \draw[thick] (6.5,-3) -- (7,-3);
        \draw (7,0) node[above] {$\varphi_{1}$};
        \draw (7,-3) node[below] {$\tilde{\varphi}_{1}$};
        \draw (7,0) to [capacitor={info=$C_g$}] (7,-3);
        \paralnode{(7,0)};
        \paralnode{(7,-3)};
        \draw (10,0) node[above] {$\varphi_{0}$};
        \draw (10,-3) node[below] {$\tilde{\varphi}_{0}$};
        \begin{scope}[shift={(10,0)}]
            \draw[thick] (0,0) -- (0,-0.8);
            \draw[thick] (-0.8,-0.8) -- (0.8,-0.8);
            \draw (0.8,-0.8) to [capacitor={info=$C_J$}] (0.8,-2.2);
            \begin{scope}[shift={(-0.8,-0.8)}]
                \draw (0,0) -- (0,-1.4);
                \coordinate (center) at (0,-0.7);
                \draw ($(center) - (0.1,0.1)$) -- ($(center) + (0.1,0.1)$) node[midway, right] {$E_J$};
                \draw ($(center) - (0.1,-0.1)$) -- ($(center) + (0.1,-0.1)$);
            \end{scope}
            \draw[thick] (-0.8,-2.2) -- (0.8,-2.2);
            \draw[thick] (0,-2.2) -- (0,-3);
        \end{scope}    
    \end{tikzpicture}
\caption{The electrical circuit under consideration. $E_J$ and $C_J$ correspond to the qubit under investigation.
    In the original experimental setup~\cite{Kuzmin2025}, a SQUID is located at the right end 
    of this circuit; in our model, we replace it with an effective Josephson junction (JJ) $E_J$ with capacitance $C_J$.
    The parameters $E_l$, $C_l$, and $C_g$ describe the transmission line (bulk), in which the phase $\varphi$ propagates.}
    \label{fig:circuit_Manuchuryan}
\end{figure}
\begin{figure}
    \centering
    \begin{tikzpicture}[
    >={Stealth[length=3mm]},
    thick,
    port/.style={draw, rectangle, minimum width=0.8cm, minimum height=1.5cm, align=center, fill=lightgray},
    line/.style={draw, thick},
    wiggle/.style={
        decorate,
        decoration={snake, amplitude=1.5pt, segment length=5pt}
    },
    scale=0.8
]
\node[port] (port) at (-0.5,0) {Port};
\coordinate (leftTop) at (0, 0.3);
\coordinate (leftBottom) at (0, -0.3);
\coordinate (rightTop) at (6, 0.3);
\coordinate (rightBottom) at (6, -0.3);
\draw[line] (leftTop) -- (rightTop);
\draw[line] (leftBottom) -- (rightBottom);
\draw[thick] (6-0.2, 0.15) rectangle (6+0.2, -0.15);
\draw[thick] (6-0.2, 0.15) -- (6+0.2, -0.15);
\draw[thick] (6-0.2, -0.15) -- (6+0.2, 0.15);
\draw[thick] (6, 0.3) -- (6, 0.15);
\draw[thick] (6, -0.3) -- (6, -0.15);
\node at (6+0.2+0.5, 0) {$r(\omega)$};
\node at (-0.55, -1.2) {$\hat{S}_L$};
\draw[wiggle] (-2.5, 0.3) -- (-1.4, 0.3) node[midway, above] {$a_L$};
\draw[-{Triangle[length=2mm, width=1.5mm]}] (-1.4,0.3) -- (-1.1,0.3);
\draw[wiggle] (-1.1, -0.3) -- (-2.2, -0.3) node[midway, below] {$b_L$};
\draw[-{Triangle[length=2mm, width=1.5mm]}] (-2.2,-0.3) -- (-2.5,-0.3);
\draw[wiggle] (0.1, 0.5) -- (1.2, 0.5) node[midway, above] {$B_L$};
\draw[-{Triangle[length=2mm, width=1.5mm]}] (1.2,0.5) -- (1.5,0.5);
\draw[wiggle] (1.5, -0.5) -- (0.4, -0.5) node[midway, below] {$A_L$};
\draw[-{Triangle[length=2mm, width=1.5mm]}] (0.4,-0.5) -- (0.1,-0.5);
\draw[wiggle] (4.4, 0.5) -- (5.5, 0.5) node[midway, above] {$B_R$};
\draw[-{Triangle[length=2mm, width=1.5mm]}] (5.5,0.5) -- (5.8,0.5);
\draw[wiggle] (5.8, -0.5) -- (4.7, -0.5) node[midway, below] {$A_R$};
\draw[-{Triangle[length=2mm, width=1.5mm]}] (4.7,-0.5) -- (4.4,-0.5);
\end{tikzpicture}
\caption{The equivalent setup and plasma-wave scattering in it.
$\hat{S}_L$ is the scattering matrix of the port at the left end of the transmission line~\eqref{eq:scattering_L}. 
$r(\omega)$ is the reflection coefficient of the signal from the only qubit $E_J$.}
\label{fig:scattering_scheme}
\end{figure}
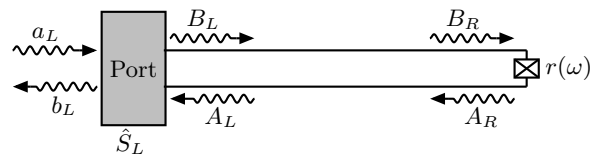

We will investigate the system depicted in Fig.~\ref{fig:circuit_Manuchuryan}: a long JJ chain terminated by a SQUID to the right and by an antenna to the left studied in experiments \cite{Kuzmin2019NP} and \cite{Kuzmin2019NPJ}. We assume the junctions of the chain to have Josephson energies equal and large enough to confine the phase difference to a minimum of the junction potential, thus, allowing us to replace them by inductances. On the other hand, we will treat the SQUID as a non-linear element. These assumptions allow us to disregard phase slips inside the chain leaving the SQUID the the only source of non-linearity in the system.

As it is widely known and will be shown below as well, in the long-wavelength limit the JJ chain functions as a transmission line. Therefore, the experimental setup is equivalent to a long transmission line with a single Josephson junction attached to its right end and terminated by a low-transparency scatterer (''port'') at the left which is schematically depicted in Fig.~\ref{fig:scattering_scheme}. In the presence of a low-transparency scatterer, the chain hosts well-defined plasma modes which acquire finite lifetimes due to non-elastic scattering off the SQUID.

Our observable of interest is the frequency-dependent reflection coefficient $S_{11}(\omega) = b_L/a_L$ relating the amplitudes of the incoming and reflected signals at the left side of the port.

\subsection{Josephson Junction Chain as a Transmission Line}
In this section we are going to construct the effective action for the qubit and the transmission line.  The experimental setup can be represented by the circuit shown in Fig.~\ref{fig:circuit_Manuchuryan}. 
Here we associate with each node (wire) its corresponding phase $\varphi_j$ for the upper line and 
$\tilde{\varphi}_j$ for the lower one. We also assume that the phase is constant along the wire.

First of all, we write the Lagrangian for this system where 
$\{\varphi_j, \tilde{\varphi}_j\}_{j=0}^N$ are dynamical variables:
\begin{equation}
    \begin{split}
        &\mathcal{L} = \frac{C_J }{8 e^2} (\dot{\tilde{\varphi}}_0 - \dot{\varphi}_0)^2
        + E_J \cos(\tilde{\varphi}_0 - \varphi_0) + \\
        &+ \sum_{j=1}^N\left[
            \frac{C_l }{8 e^2} (\dot{\varphi}_j - \dot{\varphi}_{j-1})^2 + E_l 
            \cos(\varphi_{j} - \varphi_{j-1})\right] + \\ 
            &+ \sum_{j=1}^N
            \left[\frac{C_l }{8 e^2} (\dot{\tilde{\varphi}}_j - \dot{\tilde{\varphi}}_{j-1})^2 + 
            E_l \cos(\tilde{\varphi}_j - \tilde{\varphi}_{j-1})
        \right] + \\
        &+ \sum_{j=1}^N \frac{C_g }{8 e^2} (\dot{\varphi}_j - \dot{\tilde{\varphi}}_j)^2,
    \end{split}
    \label{eq:L_discret_general}
\end{equation}
where $C_l$, $C_g$, $C_J$ stand for the capacitances  and $E_l, E_J$ stand for the Josephson energies of the junctions as shown in Fig.~\ref{fig:circuit_Manuchuryan}.


In the experiments reported in \cite{Kuzmin2019NP} and \cite{Kuzmin2019NPJ}, 
the wavelength is $\lambda \sim 10^{-4}$~m, the size of one Josephson junction is $l \sim 10^{-6}$~m, 
and the length of both two parallel chains of Josephson junctions is $L \sim 10^{-2}$~m.
A comparison of $l$ and $L$ indicates that we can use the continuum limit by taking $l \rightarrow 0$. Conveniently, in this limit the long-wavelength modes of the system are insensitive to the small differences between the Josephson energies of the chain's junctions which justifies setting them equal. Neglecting the phase slips inside the chain, we set $\varphi_j - \varphi_{j+1} \approx l \partial_x \varphi $ and obtain the following long-wavelength Lagrangian:
\begin{equation}
\begin{split}
        &\mathcal{L} = \frac{C_J }{8e^2} (\dot{\varphi}(t, 0) - \dot{\tilde{\varphi}}(t, 0))^2 + \\
        &+ E_J \cos(\varphi(t, 0) - \tilde{\varphi}(t, 0)) + \\
        &+ \frac{C}{8e^2} \int_{0}^L \mathrm d x \; 
        (\partial_t \varphi(t, x) - \partial_t \tilde{\varphi}(t, x))^2 - \\
        &- \frac{1}{4e^2 \mathcal{L}_{kin}} \int_{0}^L \mathrm d x \; \left((\partial_x \varphi(t, x))^2 + (\partial_x \tilde{\varphi}(t, x))^2\right),
\end{split}
\label{eq:Lagrangian_longwave}
\end{equation}
where $C = C_g/l$ stands for the capacitance per unit length and $\mathcal{L}_{kin} = E_l l/(2 e^2)$ stands for the line's kinetic inductance per unit length. 
Notice the gauge symmetry 
\begin{equation}
    \varphi_j(t) \rightarrow \varphi_j(t) + f(t), \quad 
    \tilde{\varphi}_j(t) \rightarrow \tilde{\varphi}_j(t) + f(t),
    \label{eq:gauge_transform}
\end{equation}
where $f(t)$ is an arbitrary function.

For further manipulations we need to transform the fields and consider
\begin{equation}
    \begin{cases}
        \bar{\varphi} = \dfrac{\varphi + \tilde{\varphi}}{2}, \\[6pt]
        \phi = \varphi - \tilde{\varphi}.
    \end{cases}
\end{equation}
After this transformation Lagrangian~\eqref{eq:Lagrangian_longwave} contains no time-derivative of 
$\bar{\varphi}(t, x)$, hence $\bar{\varphi}(t, x) \equiv \bar{\varphi}(x)$, i.e., it is constant in time and can 
be dropped from the Lagrangian. In reality, the configuration of $\bar{\varphi}(x)$ is defined by the boundary condition at $x = L$  (a dipole antenna is located here) and by the capacitive coupling of the system to the substrate. This variable is responsible for the symmetric redistribution of charge along the chain and does not produce currents or voltages between the upper and lower branches. Given that the system is electrically insulated from its surroundings and assuming the probing signal to only induce voltage on the antenna (between the branches), we may neglect the possible slow dynamics of the overall charge and safely set $\bar{\varphi}(t, x) \equiv 0$.

Finally, similarly to Refs.(\cite{ALRSZ,RSZ19}) we introduce the capacitive energy $E_C = \frac{e^2}{2C_J}$, the impedance of the transmission line $Z = \sqrt{\mathcal{L}_{kin}/C}$, the dimensionless conductance $g = R_q/Z$ (here we use the usual resistance quantum $R_q = \frac{2\pi \hbar}{e^2} \approx 25.8$ kOhm) and the velocity of the plasma mode $v = 1/\sqrt{\mathcal{L}_{kin} C}$ and obtain the Lagrangian expressed solely in terms of $\phi$:
\begin{equation}
\begin{split}
        \mathcal{L} &= \frac{1}{16 E_C} \dot{\phi}^2(t, 0)
        + E_J \cos(\phi(t, 0)) + \\
        &+ \frac{g}{16\pi} \int_{0}^L \mathrm d x \; \left[
            \frac{1}{v} \dot{\phi}^2(t, x)
            - v (\partial_x\phi(t, x))^2
        \right].
\end{split}
\label{eq:Lagrangian_longwave_reduced}
\end{equation}
This action yields the equation of motion of a propagating wave valid inside the transmission line for wavelengths $\lambda \gg l$
\begin{equation}
\partial_t^2 \phi - v^2\partial_x^2 \phi = 0.\label{eq.mot}
\end{equation}

\subsection{Scattering of Plasma Modes}
The experimental study of the circuit under consideration involves measuring the reflection of the whole system, 
since isolating the reflection of the qubit alone is difficult. The function $S_{11}(\omega)$ is used for this purpose; 
in particular, it compares the amplitudes of the outgoing and incoming modes that enter and leave the same port 
(the indices $11$ reflect this fact).

First, we assume that the scattering matrix of the port (located at $x = L$) is given by
\begin{equation}
    \hat{S}_L = \begin{pmatrix}
        r_L & t_L \\ t_L & r_L
    \end{pmatrix}
\end{equation}
where we will treat $t_L$, $r_L$ as independent of the frequency $\omega$. The subscript ``$L$'' indicates that the quantity belongs to the left boundary of the chain, i.e., $x = L$. 
The scattering at the port is lossless; therefore, $\hat{S}_L$ is unitary.

Let us define the wave amplitudes. $a_L$ and $A_L$ correspond to waves incoming to the port from the external chain 
(the waveguide) and from the transmission line, respectively, while $b_L$ and $B_L$ correspond to outgoing waves
(see Fig.~\ref{fig:scattering_scheme}). 
Thus, the scattering at the port is described by the relation
\begin{equation}
    \begin{pmatrix}
        b_L \\ B_L
    \end{pmatrix} = \hat{S}_L 
    \begin{pmatrix}
        a_L \\ A_L
    \end{pmatrix}.
    \label{eq:scattering_L}
\end{equation}

Along the JJ chain, the plasma modes propagate freely, only obtaining a phase shift of ${\omega L}{/v}$. At the right end, they undergo scattering (both elastic and non-elastic) which is described by the reflection coefficient $r(\omega)$. Defining $B_R$ as the amplitude 
incoming to the qubit and $A_R$ as the outgoing amplitude (see Fig.~\ref{fig:scattering_scheme}), 
we obtain the relation
\begin{equation}
    A_R = r(\omega) B_R. 
    \label{eq:scattering_R}
\end{equation}
Taking into account the phase shift acquired by propagating signals, we get
\begin{equation}
    A_L = A_R e^{i \frac{\omega}{v} L}, \quad B_R = B_L e^{i \frac{\omega}{v} L}
    \label{eq:scattering_phase_shift}
\end{equation}
and, finally, the desired
reflection coefficient of the whole system (Fig.~\ref{fig:S11}):
\begin{equation}
    \begin{split}
        S_{11}(\omega) &= \frac{b_L(\omega)}{a_L(\omega)} = 
        r_L + \frac{t_L^2 r(\omega) e^{2i \frac{\omega}{v} L}}{1 - r_L r(\omega) e^{2i \frac{\omega}{v} L}} = \\
        &= \frac{r_L-r(\omega)e^{2i \frac{\omega}{v} L}}{1 - r_Lr(\omega)e^{2i \frac{\omega}{v} L}}.
    \end{split}
    \label{eq:S_11}
\end{equation}
It is easy to see that all the interaction-related physics is encoded in $r(\omega)$ which will be obtained in the following sections.
\newcommand{\sympetal}[4]{
    \begin{scope}[shift={#1}, rotate=#2]
        \draw[line width=0.8pt, fill=white] (0,0) 
            .. controls (#3*0.65, -#4) and (#3, -#4*0.55) .. (#3, 0)
            coordinate[pos=0.68] (sideR)
            coordinate[pos=1.00] (tip)
            .. controls (#3, #4*0.55) and (#3*0.65, #4) .. (0,0)
            coordinate[pos=0.32] (sideL);
    \end{scope}
}

\newcommand{\doublesympetal}[4]{
    \begin{scope}[shift={#1}, rotate=#2]
        \draw[line width=0.7pt, double, double distance=0.7pt, fill=white] (0,0) 
            .. controls (#3*0.65, -#4) and (#3, -#4*0.55) .. (#3, 0)
            coordinate[pos=0.68] (sideR)
            coordinate[pos=1.00] (tip)
            .. controls (#3, #4*0.55) and (#3*0.65, #4) .. (0,0)
            coordinate[pos=0.32] (sideL);
    \end{scope}
}

\section{Schmid-Bulgadaev transition in Keldysh diagrammatic technique}\label{sec:self_energy}

\subsection{Transmission Line as an Ohmic Dissipative Environment}
In order to find the reflection coefficient of the SQUID $r(\omega)$, it is convenient to integrate out the bulk degrees of freedom of the transmission line leaving the action dependent only on $\Delta\varphi(t,x=0)$. For this purpose, we factorize the phase into the transmission line variables with zero boundary conditions 
$\chi(t, x)$ and the qubit variable $\varphi(t)$:
\begin{equation}
    \phi(\omega, x) = \chi(\omega, x) + f(\omega, x) \varphi(\omega)
    \label{eq:factorization_varphi}
\end{equation}
where $f(\omega, x)$ is an arbitrary function satisfying $f(\omega, 0) = 1$, $f(\omega, L) = 0$. 
However, for the subsequent functional integration over $\chi$ it is convenient to choose $f(\omega, x)$ 
satisfying the equation of motion in the transmission line: 
\begin{equation}
    f(\omega, x) = \frac{\sin\left(\frac{\omega}{v}(L - x)\right)}{\sin\left(\frac{\omega}{v} L\right)}.
    \label{eq:f}
\end{equation}
Substituting Eq.~\eqref{eq:factorization_varphi} into the Lagrangian \eqref{eq:Lagrangian_longwave_reduced} and 
performing integration by parts we obtain the action with factorized degrees of freedom 
($\varphi \equiv \varphi(\omega)$)
\begin{equation}
    \begin{split}
        &S[\chi, \varphi] =
        S_q[\varphi] + S_w[\chi] + S_{\text{int}}[\varphi], \\
        &S_q[\varphi] = \int_{\mathbb{R}} \mathrm d t \; \left[
            \frac{1}{16 E_C} \dot{\varphi}^2(t) + E_J \cos(\varphi(t))
        \right], \\
        &S_w[\chi] = \frac{g}{16 \pi} \int_{\mathbb{R}} \mathrm d t \int_0^L \mathrm d x \;
        \left[
            \frac{1}{v}(\partial_t \chi(t, x))^2 - v (\partial_x \chi(t, x))^2
        \right], \\
        &S_{wq}[\varphi] = -
        \frac{g}{16\pi}\int_{\mathbb{R}} \frac{\mathrm d \omega}{2\pi} \;
        |\varphi(\omega)|^2 \left[ f(-\omega, x) \partial_x f(\omega, x) \right]\bigg|_{x=0}^{L}.
    \end{split}
    \label{eq:separated_form}
\end{equation}

The transmission line plays the role of an environment that transfer energy to and from 
the considered qubit, so physically it can be introduced as an ohmic resistance. To show this, we need to 
perform functional integration over the bulk variable $\chi$:
\begin{equation}
    e^{i S_{\text{eff}}[\varphi]} = \int \mathcal{D}[\chi] \, e^{i S[\chi, \varphi]}.
    \label{eq:func_int}
\end{equation}
Action~\eqref{eq:separated_form} is factorized and quadratic in $\chi$, so the functional integration is trivial here. 
Substituting Eq.~\eqref{eq:f} into the expression for $S_{\text{int}}[\varphi]$ in Eq.~\eqref{eq:separated_form} we explicitly obtain 
\begin{equation}
    \begin{split}
        &\int_{\mathbb{R}} \frac{\mathrm d \omega}{2\pi} \; |\varphi(\omega)|^2 
        \left[ f(-\omega, x) \partial_x f(\omega, x) \right]\bigg|_{x=0}^{L} = \\
        &= \int_{\mathbb{R}} \frac{\mathrm d \omega}{2\pi} \; |\varphi(\omega)|^2 \cot\left(\frac{(\omega + i0)}{v}L\right),
    \end{split}
\end{equation}
so taking the limit $L \rightarrow +\infty$ we obtain
\begin{equation}
    \begin{split}
        S_{\text{diss}}[\varphi]
        &=
         i \frac{g}{16 \pi} 
        \int_{\mathbb{R}} \frac{\mathrm{d} \omega}{2\pi} \; \omega \, \varphi(-\omega) \varphi(\omega),
    \end{split}
    \label{eq:S_diss}
\end{equation}
where we adopt the retarded (advanced) regularization of the frequencies 
$\omega \rightarrow \omega \pm i \varepsilon$, $\varepsilon > 0$, and use the fact that 
$\cot\left(\frac{\omega \pm i \varepsilon}{v} L\right) \underset{L \rightarrow +\infty}{\rightarrow} \mp i$. 
This part of the action is imaginary and the integrand contains $\omega$ to the first power, 
which indicates dissipation with linear spectral density. 
Moreover, the presence of this term proves that the transmission line can be regarded as a resistance for frequencies $\omega \gg v/L$. 

To sum up, the effective action for the qubit is that of a resistively shunted Josephson junction. In the retarded regularization it is given by
\begin{equation}
    \begin{split}
        S_{\text{eff}}[\varphi] &=
        \int_{\mathbb{R}} \frac{\mathrm d \omega}{2\pi} \; \left(
            \frac{1}{16 E_C}\omega^2
            + \frac{g}{16 \pi} i\omega
        \right) |\varphi(\omega)|^2 + \\
        &+ E_J \int_{\mathbb{R}} \mathrm d t \; \cos(\varphi(t)) \equiv \\
        &\equiv S_0[\varphi] + S_{\text{int}}[\varphi]
    \end{split}
    \label{eq:S_eff}
\end{equation}
with an implied infrared cut-off $\pi v/L$.

It is well-known (see~\cite{Schmid1983},\cite{SchonZaikin1990}) 
that the system defined by the action given in Eq.(\ref{eq:S_eff}) exhibits the so called 
Schmid-Bulgadaev quantum phase transition between a superconducting state an insulating state at $g = 4$. 
In contrast to \cite{RSZ19} where the transition occurs at $g=2$, 
here a long transmission line can be considered semi-infinite and hence, contributes only 
half of the action shifting the transition point to $g=4$ as in \cite{RSZJSNM20}.

\subsection{Keldysh Action}
In order to capture the dynamic response of the system on both sides of  the Schmid-Bulgadaev transition, 
we will employ Keldysh diagrammatic technique. Introducing the Keldysh time contour and defining phase 
variables $\varphi_{F,B}$ on its forward and backward branches, we introduce the Keldysh action as 
$S = S_{\text{eff}}[\varphi_F] -S_\text{eff}[\varphi_B]$ (see~\cite{Kamenev2023}). 
After performing the Keldysh rotation according to
\begin{equation}
    \begin{cases}
        \varphi_{\rm cl} = \dfrac{\varphi_F + \varphi_B}{2}, \\[6pt]
        \varphi_q = \dfrac{\varphi_F - \varphi_B}{2},
    \end{cases}
    \label{eq:varphi_basis}
\end{equation}
the action takes form $S[\varphi] = S_0 + S_{\text{int}}$ with
\begin{equation}
    \begin{split}
        &S_0[\varphi] = \frac{1}{2}\sum\limits_{a,b = cl, q}\int_{\mathbb{R}} \frac{d\omega}{2\pi}\varphi_a(-\omega) \left( G_0^{-1}(\omega)\right)_{ab} \varphi_b(\omega),
         \\
        &S_{\text{int}}[\varphi] = -2E_J \int_{\mathbb{R}} \mathrm d t \; \sin(\varphi_{\rm cl}(t)) \sin(\varphi_q(t)).
    \end{split}
    \label{eq:S_eff}
\end{equation}
The free part of the action $S_0[\varphi]$ involves the inverse bare Green's function $\hat{G}_0$ given by
\begin{equation}
\hat{G} = \begin{pmatrix}
G_0^K & G_0^R \\
G_0^A & 0
\end{pmatrix}
\end{equation}
with
\begin{align}
G_0^R(\omega) &= \frac{4\pi}{g} \frac{1}{i\omega + \frac{\pi}{g E_C}\omega^2},\nonumber\\
G_0^A(\omega) & = \left[ G_0^R(\omega)\right]^\dag = G_0^R(-\omega)\nonumber\\
G_0^K(\omega) & = \coth\left(\frac{\omega}{2 T}\right)\left(G_0^R(\omega) - G_0^A(\omega)\right).
\end{align}
As it will be shown later, the reflection coefficient $r(\omega)$ can be expressed in terms of the full Green's function of phase at the boundary
\begin{equation}
    \langle \varphi_a(t) \varphi_b(t') \rangle = i G_{ab}(t, t'),
\end{equation}
or more precisely, in terms of the retarded component of the qubit self-energy $\hat{\Sigma}$ defined as
\begin{equation}
\left(\hat{G}_0^{-1} - \hat{\Sigma}\right) \hat{G} = \hat{1}.\label{self_energy_def}
\end{equation}

\subsection{Qubit Self-Energy for $E_J/E_C \ll 1$}
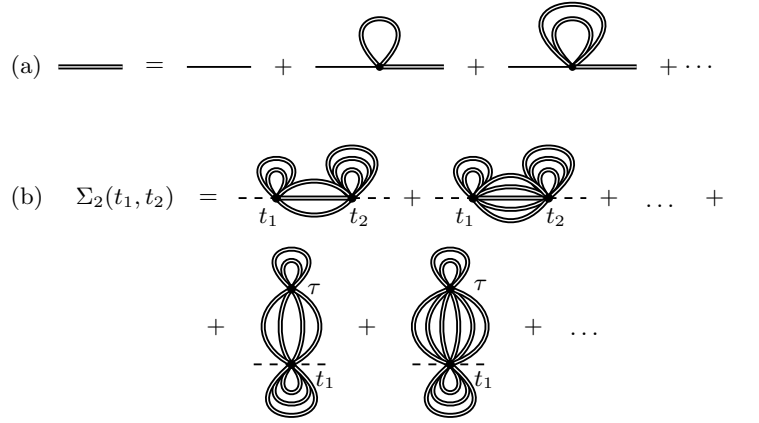
\begin{figure}
    \centering

    \begin{tikzpicture}[scale=0.85]
        \node at (-0.5, 0) {(a)};
        \draw[line width=0.7pt, double, double distance=0.5pt, fill=white] (0, 0) -- (1, 0);
        \node at (1.5, 0) {$=$};
        \draw[line width=0.7pt] (2, 0) -- (3, 0);
        \node at (3.5, 0) {$+$};
        \draw[line width=0.7pt] (4, 0) -- (5, 0);
        \begin{scope}[shift={(5, 0)}]
            \draw[line width=0.7pt, double, double distance=0.5pt, fill=white] (0,0) .. controls (1,1) and (-1,1) .. (0,0);
        \end{scope}
        \draw[line width=0.7pt, double, double distance=0.5pt, fill=white] (5, 0) -- (6, 0);
        \filldraw[black] (5,0) circle (1.5pt);
        \node at (6.5, 0) {$+$};
        \draw[line width=0.7pt] (7, 0) -- (8, 0);
        \begin{scope}[shift={(8, 0)}]
            \draw[line width=0.7pt, double, double distance=0.5pt] (0,0) .. controls (1,1) and (-1,1) .. (0,0);
            \draw[line width=0.7pt, double, double distance=0.5pt] (0,0) .. controls (1.7,1.3) and (-1.7,1.3) .. (0,0);
        \end{scope}
        \draw[line width=0.7pt, double, double distance=0.5pt, fill=white] (8, 0) -- (9, 0);
        \filldraw[black] (8,0) circle (1.5pt);
        \node at (9.5, 0) {$+$};
        \node at (10, 0) {$\dots$};
    \end{tikzpicture}

    \vspace{0.5cm}    

    \begin{tikzpicture}[scale=1]
        \node at (-1.8, 0) {(b)};
        \node at (-0.5, 0) {$\Sigma_2(t_1, t_2)$};
        \node at (0.6, 0) {$=$};
        \begin{scope}[shift={(1.5, 0)}]
            \draw[line width=0.7pt, dashed] (-0.5, 0) -- (0, 0);
            \draw[line width=0.7pt, dashed] (1, 0) -- (1.5, 0);
            \draw[line width=0.7pt, double, double distance=0.5pt] (0,0) .. controls (-0.4, 0.5) and (0.4, 0.5) .. (0,0);
            \draw[line width=0.7pt, double, double distance=0.5pt] (0,0) .. controls (-0.8, 0.7) and (0.8, 0.7) .. (0,0);
            \begin{scope}[shift={(1, 0)}]
                \draw[line width=0.7pt, double, double distance=0.5pt] (0,0) .. controls (-0.4, 0.5) and (0.4, 0.5) .. (0,0);
                \draw[line width=0.7pt, double, double distance=0.5pt] (0,0) .. controls (-0.8, 0.7) and (0.8, 0.7) .. (0,0);
                \draw[line width=0.7pt, double, double distance=0.5pt] (0,0) .. controls (-1.1, 0.9) and (1.1, 0.9) .. (0,0);
            \end{scope}
            \draw[line width=0.7pt, double, double distance=0.5pt] (0,0) -- (1,0);
            \draw[line width=0.7pt, double, double distance=0.5pt] (0,0) .. controls (0.25, 0.3) and (0.75, 0.3) .. (1,0)
            .. controls (0.75, -0.3) and (0.25, -0.3) .. (0, 0);
            \filldraw[black] (0,0) circle (1.5pt);
            \node at (-0.1, -0.25) {$t_1$};
            \filldraw[black] (1,0) circle (1.5pt);
            \node at (1.1, -0.25) {$t_2$};
        \end{scope}
        \node at (3.3, 0) {$+$};
        \begin{scope}[shift={(4.1, 0)}]
            \draw[line width=0.7pt, dashed] (-0.5, 0) -- (0, 0);
            \draw[line width=0.7pt, dashed] (1, 0) -- (1.5, 0);
            \draw[line width=0.7pt, double, double distance=0.5pt] (0,0) .. controls (-0.4, 0.5) and (0.4, 0.5) .. (0,0);
            \draw[line width=0.7pt, double, double distance=0.5pt] (0,0) .. controls (-0.8, 0.7) and (0.8, 0.7) .. (0,0);
            \begin{scope}[shift={(1, 0)}]
                \draw[line width=0.7pt, double, double distance=0.5pt] (0,0) .. controls (-0.4, 0.5) and (0.4, 0.5) .. (0,0);
                \draw[line width=0.7pt, double, double distance=0.5pt] (0,0) .. controls (-0.8, 0.7) and (0.8, 0.7) .. (0,0);
                \draw[line width=0.7pt, double, double distance=0.5pt] (0,0) .. controls (-1.1, 0.9) and (1.1, 0.9) .. (0,0);
            \end{scope}
            \draw[line width=0.7pt, double, double distance=0.5pt] (0,0) -- (1,0);
            \draw[line width=0.7pt, double, double distance=0.5pt] (0,0) .. controls (0.3, 0.2) and (0.7, 0.2) .. (1,0)
            .. controls (0.7, -0.2) and (0.3, -0.2) .. (0, 0);
            \draw[line width=0.7pt, double, double distance=0.5pt] (0,0) .. controls (0.3, 0.4) and (0.7, 0.4) .. (1,0)
            .. controls (0.7, -0.4) and (0.3, -0.4) .. (0, 0);
            \filldraw[black] (0,0) circle (1.5pt);
            \node at (-0.1, -0.25) {$t_1$};
            \filldraw[black] (1,0) circle (1.5pt);
            \node at (1.1, -0.25) {$t_2$};
        \end{scope}
        \node at (5.9, 0) {$+$};
        \node at (6.6, -0.1) {$\dots$};
        \node at (7.3, 0) {$+$};
        \begin{scope}[shift={(0.7,-1.7)}]
            \node at (0,0) {$+$};
            \begin{scope}[shift={(1,-0.5)}]
                \draw[line width=0.7pt, double, double distance=0.5pt] (0,1) .. controls (-0.4, 1.5) and (0.4, 1.5) .. (0,1);
                \draw[line width=0.7pt, double, double distance=0.5pt] (0,1) .. controls (-0.8, 1.7) and (0.8, 1.7) .. (0,1);
                \begin{scope}[shift={(0, 0)}]
                    \draw[line width=0.7pt, double, double distance=0.5pt] (0,0) .. controls (-0.4, -0.5) and (0.4, -0.5) .. (0,0);
                    \draw[line width=0.7pt, double, double distance=0.5pt] (0,0) .. controls (-0.8, -0.7) and (0.8, -0.7) .. (0,0);
                    \draw[line width=0.7pt, double, double distance=0.5pt] (0,0) .. controls (-1.1, -0.9) and (1.1, -0.9) .. (0,0);
                \end{scope}
                \draw[line width=0.7pt, double, double distance=0.5pt] (0,0) .. controls (0.2,0.25) and (0.2,0.75) .. (0,1)
                .. controls (-0.2,0.75) and (-0.2,0.25) .. (0,0);
                \draw[line width=0.7pt, double, double distance=0.5pt] (0,0) .. controls (0.5,0.25) and (0.5,0.75) .. (0,1)
                .. controls (-0.5,0.75) and (-0.5,0.25) .. (0,0);
                \filldraw[black] (0,0) circle (1.5pt);
                \node at (0.45, -0.2) {$t_1$};
                \filldraw[black] (0,1) circle (1.5pt);
                \node at (0.3, 1) {$\tau$};
                \draw[line width=0.7pt, dashed] (-0.5, 0) -- (0.5, 0);
            \end{scope}
            \node at (2, 0) {$+$};
            \begin{scope}[shift={(3.1,-0.5)}]
                \draw[line width=0.7pt, double, double distance=0.5pt] (0,1) .. controls (-0.4, 1.5) and (0.4, 1.5) .. (0,1);
                \draw[line width=0.7pt, double, double distance=0.5pt] (0,1) .. controls (-0.8, 1.7) and (0.8, 1.7) .. (0,1);
                \begin{scope}[shift={(0, 0)}]
                    \draw[line width=0.7pt, double, double distance=0.5pt] (0,0) .. controls (-0.4, -0.5) and (0.4, -0.5) .. (0,0);
                    \draw[line width=0.7pt, double, double distance=0.5pt] (0,0) .. controls (-0.8, -0.7) and (0.8, -0.7) .. (0,0);
                    \draw[line width=0.7pt, double, double distance=0.5pt] (0,0) .. controls (-1.1, -0.9) and (1.1, -0.9) .. (0,0);
                \end{scope}
                \draw[line width=0.7pt, double, double distance=0.5pt] (0,0) .. controls (0.15,0.3) and (0.15,0.7) .. (0,1)
                .. controls (-0.15,0.7) and (-0.15,0.3) .. (0,0);
                \draw[line width=0.7pt, double, double distance=0.5pt] (0,0) .. controls (0.4,0.3) and (0.4,0.7) .. (0,1)
                .. controls (-0.4,0.7) and (-0.4,0.3) .. (0,0);
                \draw[line width=0.7pt, double, double distance=0.5pt] (0,0) .. controls (0.65,0.2) and (0.65,0.8) .. (0,1)
                .. controls (-0.65,0.8) and (-0.65,0.2) .. (0,0);
                \filldraw[black] (0,0) circle (1.5pt);
                \node at (0.45, -0.2) {$t_1$};
                \filldraw[black] (0,1) circle (1.5pt);
                \node at (0.4, 1.05) {$\tau$};
                \draw[line width=0.7pt, dashed] (-0.5, 0) -- (0.5, 0);
            \end{scope}
            \node at (4.2,0) {$+$};
            \node at (4.9, -0.1) {$\dots$};
        \end{scope}
    \end{tikzpicture}
    
\caption{Self-energy diagrams in the diagrammatic technique. 
    (a) Dyson equation~\eqref{resummation_procedure} for the dressed Green's function $G_m$, which is 
    depicted by double lines. 
    Single lines correspond to the bare Green's function $G_m^{(0)}$.
    (b) Dressed diagrams of the second-order self-energies~\eqref{eq:sigma_2}.
    There is a summation over all possible numbers of loops in $t_1$, $t_2$, $\tau$, which determines the dressed $\Sigma_1$.
    The first row contains terms with only odd numbers of lines between $t_1$ and $t_2$, while 
    the second row contains only even numbers. 
    }
    \label{fig:Self-energy}
\end{figure}

For small values of $E_J/E_C$ one may employ the perturbation theory. Up to the second order, the corrections to the Green's function are given by the diagrams depicted in Fig.~(\ref{fig:Self-energy}). Their respective expressions are given by
\begin{equation}
\Sigma_1^R(\omega) = 2E_J {\rm e}^{\textstyle -\frac{i}{2}G_0^K(t = 0)}\text{ and}\label{eq:sigma_1}
\end{equation}
\begin{equation}
    \begin{split}
        \Sigma_2^R(\omega) &= 2E_J^2 \left[
        S_-(\omega) - S_-(0)
        \right] + \\
        &+2 E_J^2 \left[S_+(\omega) + S_+(0)\right] - \\
        &-4 E_J^2 G_0^R(\omega) {\rm e}^{\textstyle {-i}G_0^K(t = 0)},
    \end{split}
    \label{eq:sigma_2}
\end{equation}
where
\begin{equation}
S_{\pm}(t) = {\rm e}^{\displaystyle -i \bigl(G_0^K(0) \pm G_0^K(t) \bigr)}\sin{ G_0^R(t)}.
\label{eq:S_pm}
\end{equation}
For more details, see Appendix~\ref{app:green-function_self-energy}. 

The two contributions to the self-energy encompass different physical processes. 
$\Sigma_1$ encodes elastic scattering off the cosine potential of the SQUID within the 
Born approximation. $\Sigma_2$, in addition to an elastic part, 
contains non-elastic processes and thus, leads to decay of each individual mode in the spirit of Fermi golden rule.

Without further adjustments, $\Sigma_1 = 0$ as $G_0^K(t)$ diverges as $t \rightarrow 0$. 
This result is valid in the insulating regime $g < 4$ at $L \rightarrow \infty$, however, 
a more careful diagrammatic resummation is needed to capture the physics of the superconducting phase. 
An additional set of loop-type diagrams is incorporated into our diagrammatic series if one replaces 
$G_0$ with the dressed Green's function found as a solution of a self-consistent equation 
(see Fig.~\ref{fig:Self-energy}(a))
\begin{equation}
\hat{G}_m = \left(
\hat{G}_0^{-1} - \hat{\Sigma}_1
\right)^{-1}.\label{resummation_procedure}
\end{equation}

The independence of $\Sigma_1$ from $\omega$ allows us to obtain a closed-form equation on $\Sigma_1$ which, remarkably, reproduces the  result of the self-consistent variational approach widely used in studies of Schmid-Bulgadaev transition. Putting
\begin{equation}
G_m^R(\omega) = \frac{4\pi}{g} \frac{1}{i\omega + \omega^2/\Lambda - m}
\end{equation}
where $\Lambda = \frac{g E_C}{\pi}$ represents the effective ultraviolet cutoff and $m = 4\pi\Sigma_1/g$ --  the emergent mass, plugging it into Eq.~\eqref{eq:sigma_1}, we get the self-consistent equation 
\begin{equation}
\frac{m g}{4\pi} = 2E_J {\rm e}^{\textstyle -\frac{i}{2}G_m(0)},
\end{equation}
which in the limit $T \ll m$, $L \rightarrow \infty$ yields a power-law dependence of $m$ on $E_J$,
\begin{equation}
m = \left\{
\begin{matrix}
0 & ,\ g <4\\
\left(\frac{8\pi E_J}{g}\right)^\frac{g}{g-4}\Lambda^{-\frac{4}{g-4}} & ,\ g > 4. \end{matrix}\right.\label{m_dependence}
\end{equation}
It is noteworthy that the formula above exactly reproduces the result of the self-consistent variational 
approach and is in complete agreement with the renormalization group picture \cite{bulgadaev1984phase, guinea1985diffusion}, 
therefore, this expression maybe thought of as the equation for the renormalized Josephson energy $\tilde{E}_J$
\begin{equation}
\tilde{E}_J \simeq E_J{\rm e}^{\textstyle -\frac{i}{2}G_m(0)} = \frac{mg}{8\pi} = E_J \left(\frac{8 \pi^2 E_J}{g^2E_C}\right)^{\frac{4}{g -4}}\label{renormalized_E_J}.
\end{equation}
Now the physics of the Schmid-Bulgadaev transition is clear: in the insulating state for $g <4$ phase fluctuations are strong enough to completely suppress the binding potential of the cosine leaving the phase difference across the SQUID to fluctuate freely. In the superconducting phase ($g >4$), the potential survives the fluctuation renormalization, and the phase difference is confined to one of its minima. The emergent mass is effectively the order parameter of our system and its emergence signalizes a transition to a non-perturbative regime where the lowest-order perturbation theory ceases to ne applicable.
It is also worth mentioning that in a finite system, the mass will appear even for $g < 4$, although in this case it will decay with the system size as $m \propto \left(\frac{L E_C}{v}\right)^{-4/g}$.

Now let us examine $\Sigma_2$ which must also be calculated using $G_m$. For $\omega > m$ at low temperatures we have a power-law dependence of its $\omega$-dependent part responsible for the excitation lifetime,
\begin{equation}
\Sigma_2(\omega) = C(g)E_J^2 E_C^{\ts -\frac{8}{g}} |\omega|^{\ts \frac{8}{g}-1}\label{Sigma_2}
\end{equation}
with a $g$-dependent dimensionless factor $C(g)$ provided in Appendix~\ref{app:green-function_self-energy}. At lower frequencies $\omega < m$ within this approximation $\Im \Sigma_2(\omega) \propto -\omega$. 

Comparing $\Sigma_1 \propto E_J \left(\frac{E_J}{E_C}\right)^\frac{4}{g-4}$ and $\Sigma_2$, 
we find that at $g = 8$ they scale similarly. At $g > 8$ $\Sigma_2$ is a higher power in $E_J$ 
which signals that in this range it is indeed a smaller correction to $\Sigma_1$ 
while for $4 <g <8$ the situation is opposite. It is also easy to note that at $\omega\sim m$ 
we have $\Sigma_2 \sim \Sigma_1$, therefore, the low-frequency regime cannot be fully captured by perturbation theory, 
even with the resummation. Nevertheless, one may expect that $\Im \Sigma_2$ decays to zero 
as $\omega \rightarrow 0$ both from the resulting dependence within our approximation 
(see Eq.~\eqref{eq:Sigma_2_low_frequency}) and the result of the analysis in the dual regime 
$E_J \gg E_C$ which will be worked out below. A detailed analysis carried out in 
Appendix~\ref{app:green-function_self-energy} shows that at $g >8$ the dominant elastic contribution 
comes from $\Sigma_1$ while at $4 < g <8$ both $\Sigma_1$ and $\Sigma_2$ should be kept.

Finally, the Green's function in the presence of the SQUID is given by
\begin{equation}
        G^R(\omega) = \frac{1}{\frac{\omega^2}{4E_C} + i \frac{g}{4\pi} \omega - \Sigma(\omega)}.
    \label{eq:GR_exact}
\end{equation}

\subsection{Qubit Self-Energy for $E_J/E_C \gg 1$}\label{subsec:strong_coupling}

The calculations above are relevant only for the perturbative case $E_J/E_C \ll 1$.  
However, we can extend the results of that section to the opposite limit $E_J/E_C \gg 1$ 
by employing the duality between the phase and charge representations of our system \cite{SchonZaikin1990}. 
Indeed, in the dual limit it is convenient to describe the system by charge variable 
(canonical conjugated to phase $\varphi$) $\chi = \frac{\dot{\varphi}}{8 E_C}$ \cite{Schmid1983}, so after rewriting 
the partition function in expansion on instanton series classical action 
for charge variable is written as 
\begin{equation}
    \begin{split}
        S_{0}^{\text{dual}}[\chi] &= \int_{\mathbb{R}} \frac{\mathrm d \omega}{2\pi} \; 
        \left(
            \frac{\omega^2}{16 E_C} + \frac{16}{\pi g} i\omega
        \right)|\chi(\omega)|^2, \\
        S_{\text{int}}^{\text{dual}}[\chi] &= \Delta_0 \int_{\mathbb{R}} \mathrm d t \; \cos(\chi(t)).
    \end{split}
\end{equation}
Moreover, here we should apply by the Green's function corresponded to charge correlations 
$D_{ab}(\omega) = -i \langle \chi_a(-\omega) \chi_b(\omega)\rangle = 
- i \frac{\omega^2}{64 E_C^2} \langle \varphi_a(-\omega) \varphi_b(\omega) \rangle = \frac{\omega^2}{64 E_C^2} G_{ab}(\omega)$.
Also from rewriting the partition function in these limits it is known that duality holds here 
(left: $E_J/E_C \ll 1$, right: $E_J/E_C \gg 1$) \cite{SchonZaikin1990}:
\begin{equation}
    g \leftrightarrow \frac{16}{g}, \quad \frac{E_J}{2g E_C} \leftrightarrow \frac{\Delta_0}{\omega_0}, 
    \quad E_J \leftrightarrow \Delta_0.
    \label{eq:duality}
\end{equation}
where we introduced 
\begin{equation}
    \begin{split}
        &\Delta_0 = 16 \sqrt{\frac{E_J E_C}{\pi}} \left(\frac{E_J}{2E_C}\right)^{1/4} e^{-\sqrt{\frac{8 E_J}{E_C}}}, \\
        &\omega_0 = \sqrt{8 E_J E_C}.
    \end{split}
    \label{eq:Delta_omega}
\end{equation}

Physically, $\Delta_0$ defines the tunnel amplitude between two nearby cosine minima in the limit $E_J/E_C\gg1$ \cite{SchonZaikin1990}, 
this fact can be seen from exponential dependence of $\Delta_0$ on JJ amplitude as $E_J \rightarrow +\infty$.
Also it can be associated with a bond width because $\Delta_0$ enters the energy of ground state as 
$E_{\text{ground}}^{\text{dual}}(\chi) = - \Delta_0 \cos(\chi)$. 

There is instanton physics in the strong-coupling limit with a time width of $\sim \omega_0^{-1}$; 
hence, the new UV-cutoff parameter is $\sim \omega_0$ and, more accurately, 
from Eqs.~\eqref{eq:duality}, $2 \pi \Lambda \leftrightarrow \omega_0$.

Applying the duality rules~\eqref{eq:duality}, we readily obtain
the retarded Green's function $D^{\text{dual}, R}$ of the charge variable from~\eqref{eq:GR_exact} and the corresponding self‑energy $\Sigma^{\text{dual}}$~\eqref{eq:sigma_1}, \eqref{eq:sigma_2} in the limit 
$E_J/E_C \gg 1$: 
\begin{equation}
    D^{\text{dual}, R}(\omega) = \frac{1}{\frac{\omega^2}{4E_C} + i \frac{4}{\pi g} \omega - \Sigma^{\text{dual}}(\omega)}, 
    \label{eq:GR_exact_dual}
\end{equation}
\begin{equation}
    m^{\text{dual}} = 
    \begin{cases}
        \frac{16 E_C}{\pi g} \left(\frac{\pi^2 g\Delta_0}{\omega_0}\right)^{\frac{4}{4-g}}, \quad g < 4, \\
        0, \quad g > 4,
    \end{cases}
    \label{eq:dual_mass}
\end{equation}
\begin{equation}
    \begin{split}
        \Sigma_2^{\text{dual}}(\omega) &= 
        C^{\text{dual}}(g) E_C \frac{\Delta_0^2}{\omega_0^2} \left|\frac{\omega}{E_C}\right|^{\frac{g}{2}-1}, \; \omega > m^{\text{dual}}.
    \end{split}
    \label{eq:Sigma2_dual}
\end{equation}
The same analysis as after Eq.~\eqref{Sigma_2} can be applied with duality~\eqref{eq:duality}. 
Note that here $\Im\Sigma_2(\omega)$ goes to zero as $\omega^{g/2-1}$ at $g>4$ which is even faster than the asymptotic derived in the previous subsection.
\section{Scattering on the Qubit}

\subsection{Reflection from the Qubit}
In the previous section we obtained the retarded self-energy up to second order, which allows us to investigate the 
Schmid-Bulgadaev phase transition. In this section we obtain $r(\omega)$. 

In the presence of the SQUID, one obtains $r(\omega)$ by solving the equation of motion (\ref{eq.mot})
\begin{equation}
    \left(\partial_x^2 + \frac{\omega^2}{v^2}\right) G^R(\omega; x, x') = 
    \frac{4\pi}{g v} \delta(x - x')
    \label{eq:Green_equation}
\end{equation}
on the half-line $x, x' > 0$ with the boundary conditions ensuring the continuity of phase and current at the boundary, leaving the following system to be solved
\begin{equation}
    \begin{split}
        &\partial_x G^R(\omega;+0, x') + \\
        &+\left(
        \dfrac{\pi \omega^2}{g E_C v} - \dfrac{4\pi}{g v}\Sigma(\omega)
        \right) G^R(\omega; +0, x') = 0.
    \end{split}
    \label{eq:boundary_conditions}
\end{equation}
The boundary conditions mentioned above are necessary to ensure
$G^R(\omega; 0, 0)$ with $G^R(\omega)$ from the previous section where it was obtained.

The retarded Green's function is sought in the form 
\begin{equation}
    G^R(\omega; x, x') = \left\{
    \begin{matrix}
        A(\omega, x') e^{i \frac{\omega}{v} x}, & \ x > x^\prime \\ 
        B(\omega, x') e^{i \frac{\omega}{v} x} + C(\omega, x') e^{- i \frac{\omega}{v} x}, & \ x < x^\prime.
    \end{matrix}\right.
    \label{eq:reflection_Green_function}
\end{equation}
Using finding such a solution, we obtain the reflection coefficient 
\begin{equation}
    r(\omega) = \frac{B(\omega, x')}{C(\omega, x')} = 
    - \frac{\omega^2 - i \frac{gE_C}{\pi} \omega - 4E_C \Sigma(\omega)}
    {\omega^2 + i \frac{gE_C}{\pi} \omega - 4E_C \Sigma(\omega)}.\label{eq:reflection_coef}
\end{equation}

Equation~\eqref{eq:reflection_coef} demonstrates that $|r(\omega)|^2$ can be less than 1. This occurs because there are inelastic processes in the SQUID which yield non-zero $\Im(\Sigma(\omega)) < 0$. 

\begin{figure}[!t]
    \centering
    {
        \includegraphics[width=0.48\textwidth]{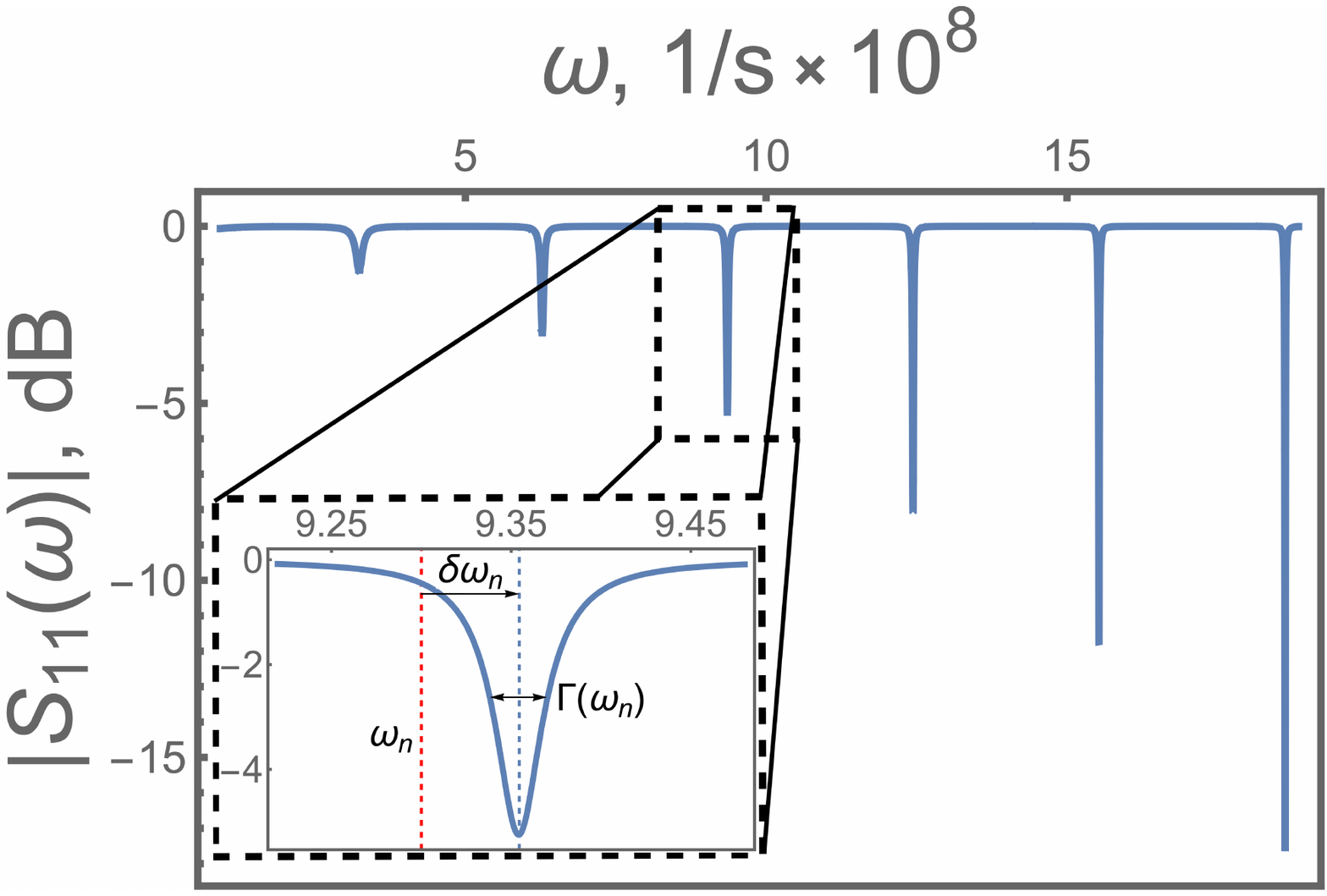}
    }
    \caption{Plot of $|S_{11}(\omega)|$ for $g = 12$ (see Eq.~\eqref{eq:S_11}). 
    It demonstrates the observable reflection in experiments \cite{Kuzmin2019NPJ}, \cite{Kuzmin2025}.
    The inset provides a zoomed plot of an isolated peak shifted by $\delta \omega_n$ with finite non-zero width $\Gamma(\omega_n)$. 
    $r_L = 0.99$, $L = 10^{-2} \text{ m}$, 
    $E_C = 300 \text{ MHz}$, $E_J/E_C = 10^{-1}$, $v = 10^6 \text{ m/s}$.}
    \label{fig:S11}
\end{figure}

Plugging Eq.(\ref{eq:reflection_coef}) into Eq.(\ref{eq:S_11}), one obtains $S_{11}(\omega)$ -- the ratio of the amplitudes of the incoming and reflected signals (see Fig.~(\ref{fig:S11})).  The ratio of their intensities is close to $1$ for all frequencies outside of the vicinities of the frequencies of individual modes of our system. Near these frequencies, a set of distinct dips located at the energies of the system's modes can be found. Below we discuss in detail the positions of these dips and their widths $\Gamma(\omega)$.

\subsection{Mode Frequency Shift}
\begin{figure}[!t]
    \centering
    {
        \includegraphics[width=0.49\textwidth]{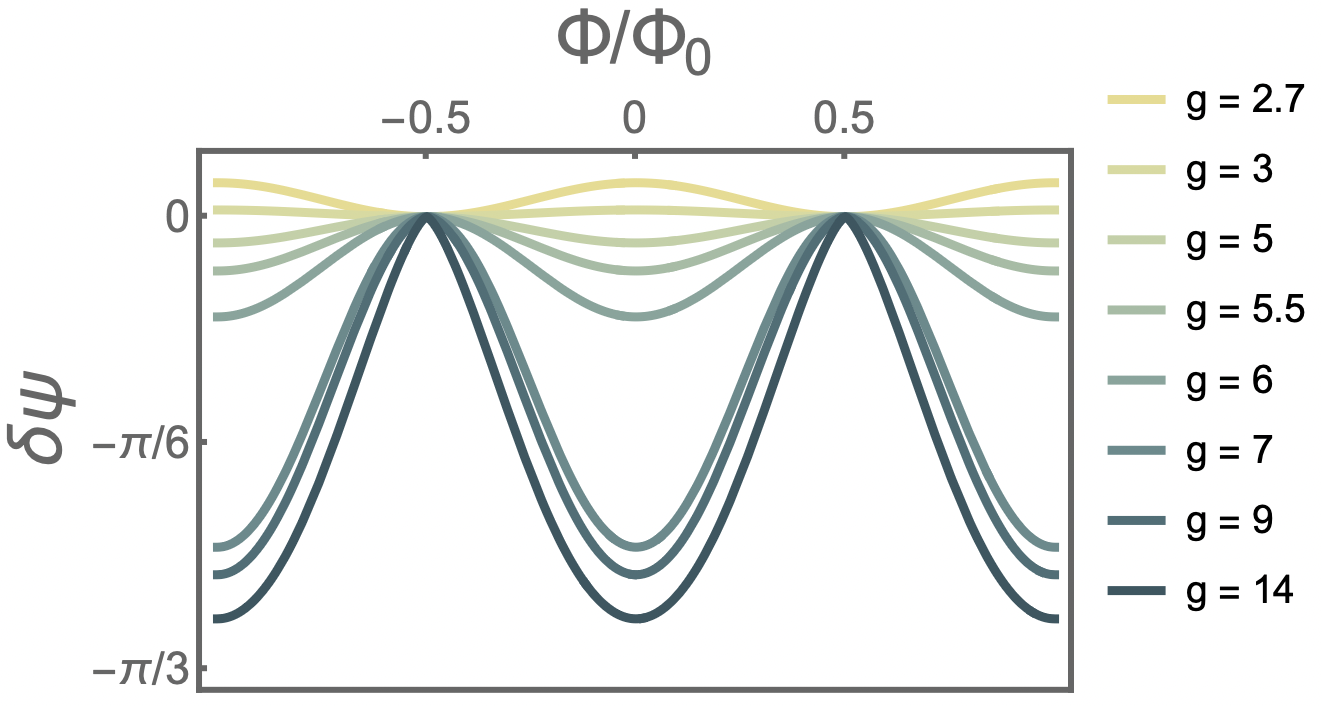}
    }
    \caption{Plots of $\delta \psi(\omega, \Phi) := \arg(r(\omega)) - \arg(r(\omega)|_{E_J = 0}$ with $\omega = E_C/10$ for different $g$.
    Plots for $g > 4$ have only mass $m$, higher orders in perturbation series are neglected.
    One can find here the special point $\frac{\Phi}{\Phi_0} = 0.5$, where the behavior of these functions 
    changes significantly with $g$ and has the power-law dependence on $|\Phi - \Phi_0/2|$.}
    \label{fig:deltapsi(Phi)}
\end{figure}
The centers of the dips are found at $\frac{2\omega_n L}{v} + \arg(r(\omega_n)) = 2\pi n$ which, taking into account that the phase of $r(\omega)$ is a slowly changing function much smaller than the length-dependent part, yields a small shift of the mode frequencies
\begin{equation}
\begin{split}
    &\delta\omega = -\frac{v}{2L}\left(\arg(r(\omega)) - \arg(r(\omega)|_{E_J = 0}\right) = \\
    &= -\frac{v}{2L}\arg\left(
\frac{\omega^2 - i \frac{gE_C}{\pi} \omega - 4E_C \Sigma(\omega)}
    {\omega^2 + i \frac{gE_C}{\pi} \omega - 4E_C \Sigma(\omega)} \cdot
\frac{\omega^2 + i \frac{gE_C}{\pi} \omega}
    {\omega^2 - i \frac{gE_C}{\pi} \omega}
\right).
\end{split}
\end{equation}
provided in Fig.(\ref{fig:deltapsi(Phi)}). 
The effective Josephson energy of the SQUID is experimentally controlled by the external flux 
$\Phi$ and for following analysis we change $ E_J \rightarrow E_{J}^{\text{eff}} = E_J\cos(\pi \Phi/\Phi_0)$, as a result, the positions of peaks are periodic functions of $\Phi$. There are two interesting features of this dependence. First, the sign of $\delta \omega$ is different in different phases. Second, at low temperatures $\delta \omega$ exhibits a power-law behavior in the vicinity of $\Phi = \Phi_0/2 + \delta\Phi$
\begin{equation}
\delta\omega \propto \left\{\begin{matrix}
- \Re(C(g)) \frac{v}{L} \frac{E_J^2}{E_C^2} \left|\frac{\omega}{E_C}\right|^{\frac{8}{g} - 2}
\left(\pi \frac{\delta \Phi}{\Phi_0}\right)^2, & g<8\\
\frac{v}{L} \frac{E_C}{\omega} \left(\frac{E_J}{E_C}\right)^{\frac{g}{g - 4}} \left|\pi\frac{\delta \Phi}{\Phi_0}\right|^{\frac{g}{g - 4}}, & g>8.
\end{matrix}
\right.\label{delta_omega}
\end{equation} 
Note that for $g > 8$ the frequency shift follows the power-law dependence 
of $\tilde{E}_J$. Thus, one one observes the change of the sign of $\delta\omega$ at the transition point 
$g = 4$ (in accordance with \cite{Kuzmin2025}) and the crossover to mass-dominated power-law behavior missed in 
\cite{Kuzmin2025}.

\subsection{Resonance Width}
\begin{figure}[!t]
    \centering
    {
        \includegraphics[width=0.49\textwidth]{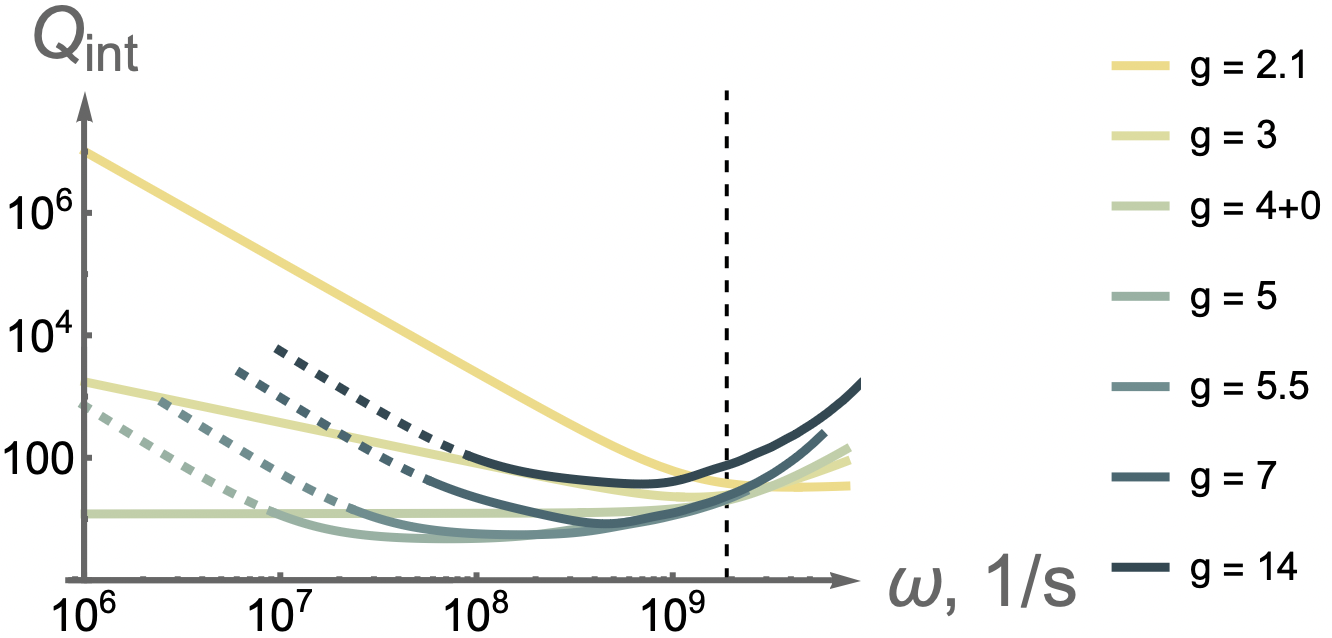}
    }
    \caption{Plots of $Q_{\text{int}}$ vs. $\omega$ in the weak-coupling limit. 
    Vertical black dashed line demonstrates the $E_C$ scale (the UV-cutoff scale). 
    Colorful lines are $Q_{\text{int}}$ plots, estimated with numerical calculations of $\Sigma$.
    Colorful dashed lines are $Q_{\text{int}}$ plots for $\omega < m$, 
    where the theory cannot guarantee the correctness of the quality factor calculated through Eq.~\eqref{eq:Q_int}.}
    \label{fig:Q_int(omega)}
\end{figure}
The shape of the dip in the vicinity of the resonance frequency $\omega_n$ can be approximated by a formula
\begin{equation}
S_{11}(\omega) \approx \frac{Q_{\text{int}}^{-1} - Q_{\text{ext}}^{-1} - {2i L(\omega - \omega_n)}/{v}
}{Q_{\text{int}}^{-1} + Q_{\text{ext}}^{-1} - {2i L(\omega - \omega_n)}/{v}}
\end{equation}
if we identify the quality factors with $Q_{\text{ext}}^{-1} = 1 - |r_L|^2$, $Q_{\text{int}}^{-1} = 1- |r(\omega)|^2$ describing the ratio of the loss of intensity throughout the scattering the port and cubit, respectively, to the incoming intensity of the signal at a frequency $\omega$. This expression differs from the well-known expression 
$S_{11}(\omega) \approx \frac{Q_{\text{int}}^{-1} - Q_{\text{ext}}^{-1} - {2i (\omega - \omega_n)}/{\omega_n}
}{Q_{\text{int}}^{-1} + Q_{\text{ext}}^{-1} - {2i (\omega - \omega_n)}/{\omega_n}}$ provided in \cite{Kuzmin2019NP} since here, we deal with the energy loss over the time the signal takes to traverse the wire from the port to the SQUID and back, instead of over one period.

\begin{equation}
    \begin{split}
        Q^{-1}_{\text{int}}(\omega)
        =-\frac{g}{\pi}\frac{\omega\, \Im\Sigma(\omega)}
        {\left(\frac{\omega^2}{4E_C} - \Re\Sigma(\omega)\right)^2 + 
        \left(
        \frac{g\omega}{4\pi} - \Im\Sigma(\omega)
        \right)^2.
        }
    \end{split}
\label{eq:Q_int}
\end{equation}
Using the obtained expression for $\Sigma$ (Eqs.(\ref{renormalized_E_J},\ref{Sigma_2})), for $E_J \ll E_C$ one recovers the high-frequency ($ \tilde{E}_J \ll \omega \ll E_C$) power-law growth
\begin{equation} 
Q_{\text{int}} \propto \left(\frac{E_C}{E_J}\right)^2\left(\frac{\omega}{E_C}\right)^{2 - 8/g}.\label{Q_small_EJ}
\end{equation}
At $\omega \ll \tilde{E}_J$ the quality factor decays with $\omega$, although, the non-perturbative behavior of the problem does not let us provide a the asymptotic formula. As a result, $Q_{\text{int}}(\omega)$ exhibits a minimum at $\omega \sim \tilde{E}_J$ for $g >4$ and $E_J \ll E_C$. In the insulating regime $g < 4$ the answer is given by Eq.(\ref{Q_small_EJ}) with a power-law decay. As a result, the Schmid-Bulgadaev transition presents itself as the change in the high-frequency behavior of the quality factor between growth and decay while the emergent mass in the superconducting phase may be estimated by the location of the minimum of $Q_{\text{int}}$.

For $E_J \gg E_C$ we are in the dual limit where Eqs.~\eqref{eq:Sigma2_dual}, \eqref{eq:dual_mass} should be used, yielding

\begin{equation}
Q_{\text{int}}(\omega) \propto \left(\frac{\omega}{\Delta_0}\right)^{2} \left(\frac{\omega}{E_C}\right)^{2 - \frac{g}{2}}, \quad \omega \gg m^{\text{dual}} \label{Q_big_EJ}
\end{equation}

Here, the non-monotonous behavior of $Q_{\text{int}}$ occurs in the insulating regime with a 
minimum at $\omega \sim m^\prime$. In the insulating regime at large frequencies we have a 
power-law growth while $Q_{\text{int}}$ decays with frequency (Eq.(\ref{Q_big_EJ})) in the superconducting regime $g >4$. Together the obtained expressions provide a recipe of detecting the Schmid-Bulgadaev transition through measuring the internal quality factors of the modes for $\omega < E_C$.

A numerical plot of $Q_{\text{int}}$ is provided in Fig.~(\ref{fig:Q_int(omega)}).
\subsection{Behavior of the System vs. Transmission Line Length}
\begin{figure}[!t]
    \centering
    {
        \includegraphics[width=0.45\textwidth]{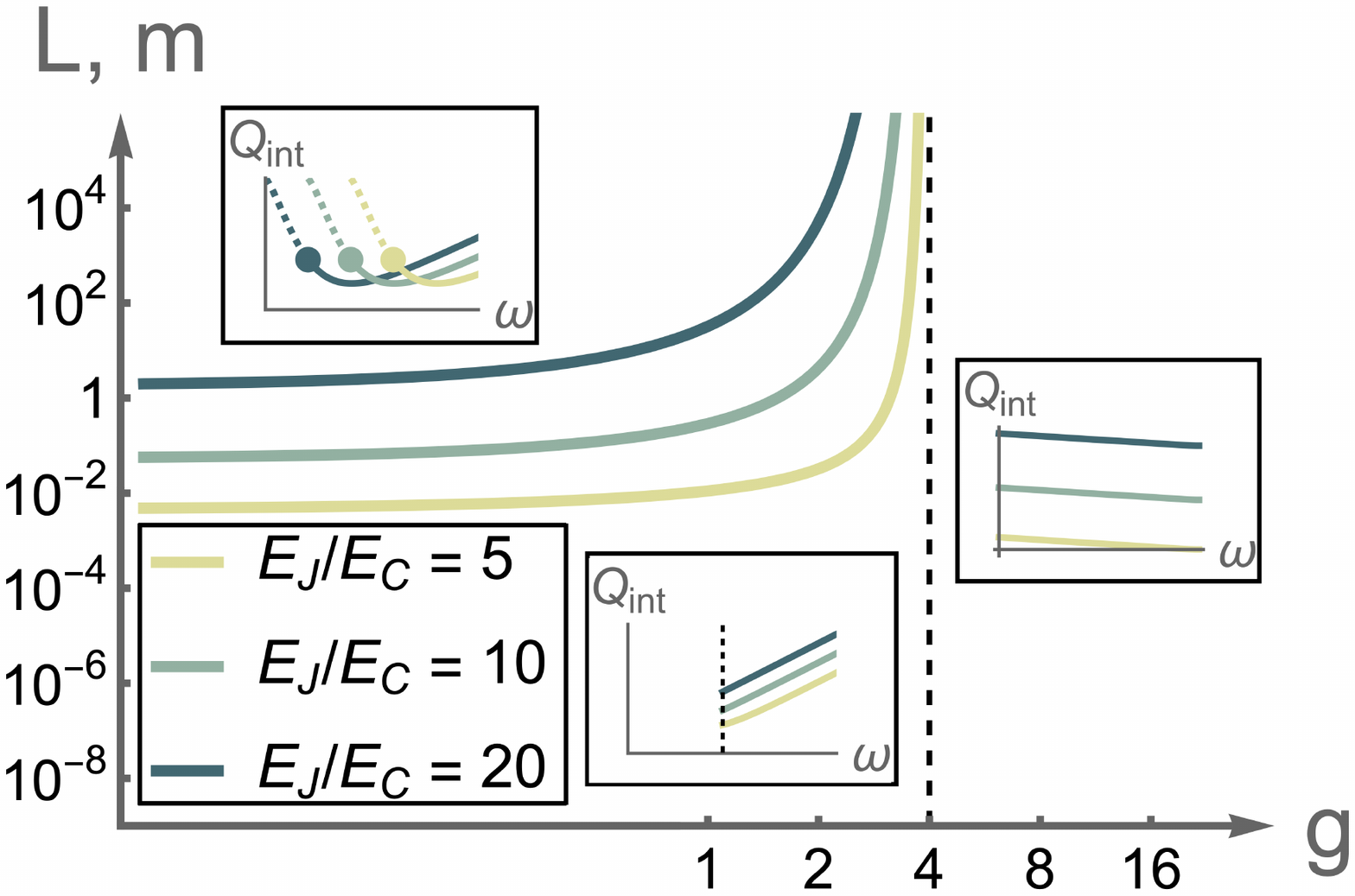}    
        \includegraphics[width=0.45\textwidth]{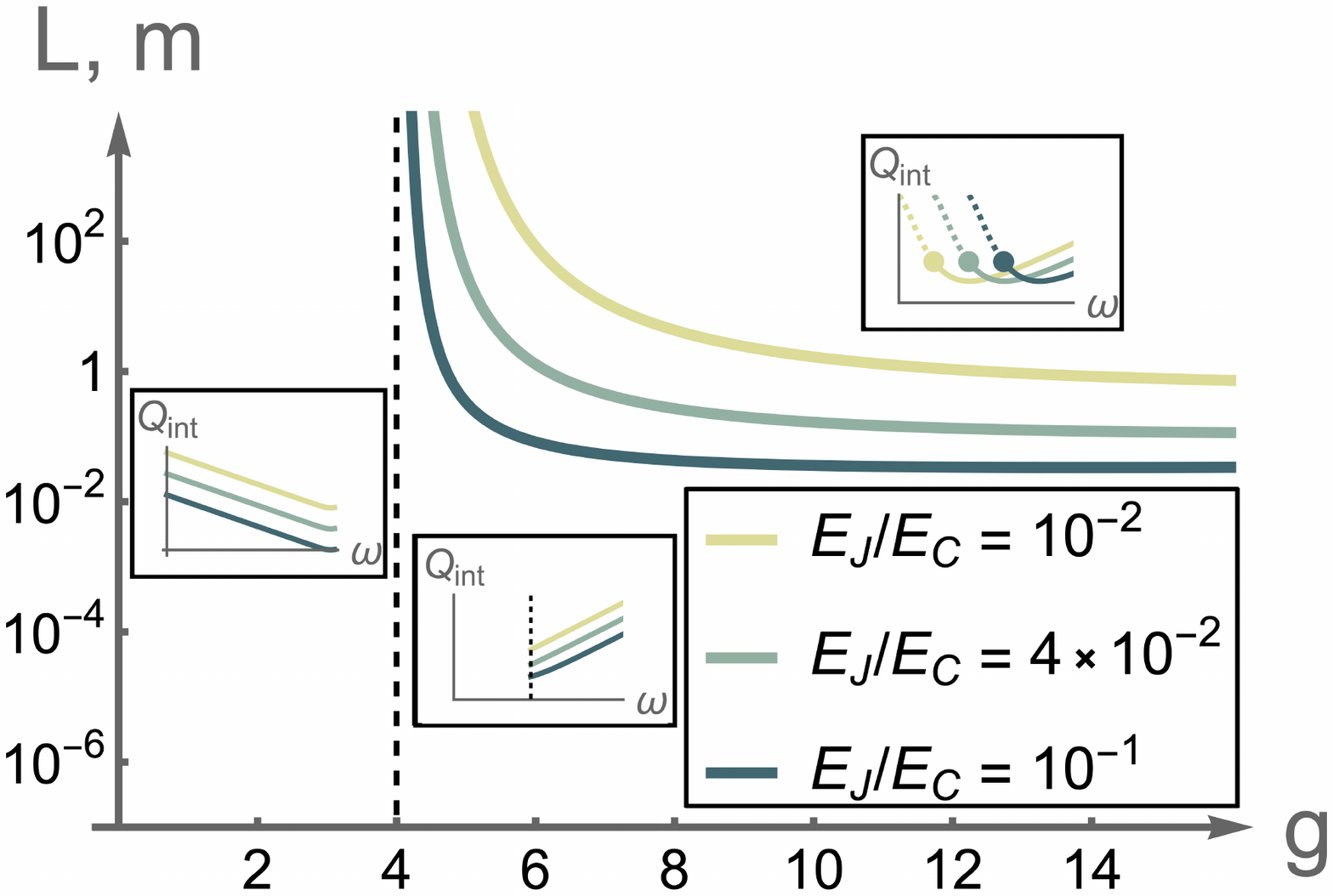}
    }
    \caption{Behavior of $Q_{\text{int}}$ versus the system parameters ($g$, $L$, $E_J/E_C$) for the strong-coupling (upper panel) and weak-coupling (lower panel) regimes. 
    The black dashed vertical line denotes the Schmid-Bulgadaev transition. 
    Solid lines in the insets illustrate the expected form of the observable quality factor~\eqref{eq:Q_int} 
    on logarithmic scales.
    The thick dots in the insets indicate the mass scale, and 
    the colored dashed lines are $Q_{\text{int}}$ plots for $\omega < m$, 
    where the theory does not guarantee the validity of the quality factor shown in Fig.~\ref{fig:Q_int(omega)}.}    
    \label{fig:L(g)}
\end{figure}
The power-law renormalization of $E_J$ requires the system to be sufficiently long, so that there are many individual modes with energies $\omega_n < m$. Thus, the obtained behavior holds as long 
as $L \gg L^*$
\footnote{In this section we present an analysis valid in the weak-coupling regime $E_J\ll E_C$. 
In the opposite limit the duality transformation provides the corresponding expressions in the 
strong-coupling regime},
\begin{equation}
    L^* = \frac{\pi v}{m} = \frac{8\pi^2 v}{g E_C}\left(\frac{E_C}{8\pi E_J}\right)^{\frac{g}{g-4}}.\label{L*}
\end{equation}
Shorter systems exhibit an $L$-dependent behavior since $\Sigma_{1}$ acquires such a dependence with the new renormalized $\tilde{E}_J$,
\begin{equation}
\tilde{E}_J(L) = E_J \left(
\frac{gE_C L}{\pi^2 v}
\right)^{-\frac{4}{g}}
\end{equation}
valid for $L < L^*$ -- in chains where the first mode lies above $m$. In such shorter wires phase fluctuations do not have enough room to fully develop, therefore the effective values of $E_J$ are larger, and one observes bigger values of $\Sigma_1$ and mode frequency shifts $\delta \omega_n$. At $g > 8$ where $\Sigma_1$ dominates, one should observe a non-analytic zero-temperature behavior in the vicinity of $\Phi = \Phi_0/2$,
\begin{equation}
\delta \omega_n\propto \frac{v}{L}\frac{E_C}{\omega_n}E_J\left(\frac{E_C L}{v}\right)^{-4/g}|\Phi/\Phi_0 - 1/2|.\label{delta_omega_small_L}
\end{equation}

In Fig.~\ref{fig:L(g)} one may find $L^*$ as a function of $g$ for different values of $E_J/E_C$ and a qualitative depiction of $Q_{\text{int}}(\omega)$ in the limits $L \ll L^*$ and $L \gg L^*$. In the experimental setup shown in \cite{Kuzmin2025}, the transmission line length was $L = 6 \text{ mm} \sim L^*$  for the weak-coupling regime, thus, the system was of intermediate length not sufficient to recover the low-energy physics of the system. Finally, it is worth noting that in the dual strong-coupling regime the exponential suppression of $\Delta_0$ (see Eqs.(\ref{eq:Delta_omega}) makes $L^*$ too large for the long-chain regime to be achieved experimentally.
\section{Discussion}

In this work we investigated the scattering properties of the system depicted in Fig.~(\ref{fig:circuit_Manuchuryan}) similar to the one used in experiment \cite{Kuzmin2025} across both superconducting ($g > 4$) and insulating ($g<4$) sides of the Schmid-Bulgadaev quantum phase transition. We employed a diagrammatic resummation summarized by Eq.~(\ref{resummation_procedure}) and Fig.~(\ref{fig:Self-energy}) which captures the non-perturbative physics of the superconducting phase introducing the emergent mass $m$ given in Eq.(\ref{m_dependence}). This emergent mass describes the effective parabolic potential of the renormalized Josephosn energy of the SQUID at the boundary which pins the phase difference to one of the minima of the SQUID's potential. Upon subsequent solution of the plasma mode scattering problem given in Eqs.~(\ref{eq:Green_equation},\ref{eq:boundary_conditions}) we obtained an analytic expression for the reflection coefficient $r(\omega)$ (see Eq.(\ref{eq:reflection_coef})) expressed in terms of the self-energy $\Sigma(\omega)$ of the qubit calculated both in the weak-coupling ($E_J\ll E_C$) and strong coupling ($E_J \gg E_C$) regime by virtue of the phase-charge duality. We investigated the dependence of the individual mode frequency shifts (see Eq.(\ref{delta_omega})) as a function of the flux through the SQUID and found that it crucially depends on the emergent mass, especially at $g > 8$ where this dependence yields a power-law low-temperature behavior, see Fig.~(\ref{fig:deltapsi(Phi)}). 

We found the internal quality factors $Q_{\text{int}}$ of individual modes of the chain (Eq.~(\ref{Q_small_EJ})) responsible for the observed width of the resonance peaks and found that $Q_{\text{int}}(\omega)$ exhibits a minimum at a frequency of order of $m$ as a result of the non-perturbative behavior of the system. Despite the strongly non-perturbative behavior of the quality factor at $\omega <m$ which prevents obtaining exact asymptotic formulas, the obtained qualitative result follows from both the weak-  and strong-coupling calculations.

We identified the critical scale $L^*$ given in Eq.~(\ref{L*}) which separates the non-perturbative limit of the long system with $L \gg L^*$ where the system exhibits $L$-independent power-law behavior and the perturbative limit of a short system where one may use unmodified perturbation theory without resummation (\ref{resummation_procedure}). Crucially, in this limit the emergent mass $m$ is even larger, growing as the system's size becomes smaller, which yields a size-dependent behavior of the frequency shifts given in Eq.~(\ref{delta_omega_small_L}) eventually leading to a dependence characteristic of a non-shunted Josephson junction. 

Together, the formulas for the frequency shifts and the internal quality factors of the plasma modes obtained in various limits constitute the main result of our work. Since the Schmid-Bulgadaev transition (as it is with but all phase transitions) manifests itself most profoundly in the low-frequency properties of the system, we primarily focused on frequencies $\omega \ll E_C$. The experiment \citep{Kuzmin2025} seemingly focused on the opposite limit of $\omega > E_C$ where the essentially only remaining signature of the transition the change of the sign of $\delta \omega_n$ between the superconducting and non-superconducting phases. In our work, we improved the theory derived in \citep{Kuzmin2025} by implementing the resummation (\ref{resummation_procedure}) which allowed us to assess the emergent mass whose appearance indicates the onset of the superconducting phase and which is known to be responsible for the ability of the junction to conduct dissipationless current \cite{kampf1987quantum}. Upon comparing it the the second-order contribution from $\Sigma_2$ we found that, especially at high $g$ or for short junctions, the mass-contribution from $\Sigma_1$ dominates even at large frequencies. This conclusion may be checked upon inspecting the dependence of these shifts on $L$ for wires with similar $g$, $E_J$, $E_C$.

Even though the phenomenon of Schmid-Bulgadaev transition in the system was investigated in \citep{Kuzmin2025}, 
the experiment was performed on chains with $L \lesssim  L^*$
in the perturbative Mermin-Wagner regime where the fluctuations of 
phase are bounded by the system's dimensions. In other words, the system's order parameter 
$m$ does not have enough room to form, leaving the mentioned signatures of the transition unobservable. 
For the experimental parameters of Ref.~\cite{Kuzmin2025}, we estimate the lowest $L^*$ 
to be $5 \times 10^{-2}\div 10^{-1} \text{ m}$ against the used $L = 6 \times 10^{-2} \text{ m}$, hence, 
raising the JJ chain's length by a single order should make the non-perturbative physics observable.

\begin{acknowledgments}

\end{acknowledgments}

\appendix
\section{Green's Functions and Expressions for Self-Energy} \label{app:green-function_self-energy}
Here we provide some necessary expressions to obtain Eqs.~\eqref{eq:sigma_1}, \eqref{eq:sigma_2}, \eqref{eq:S_pm}, \eqref{Sigma_2}.
First, we rewrite the interaction in the action~\eqref{eq:S_eff} using Eq.~\eqref{eq:varphi_basis} as 
\begin{equation}
    \begin{split}
        S_{\text{int}}[\varphi] &= E_J \int_{\mathbb{R}} \mathrm d t \; \left[\cos(\varphi_F(t)) - \cos(\varphi_B(t))\right] = \\
        &= E_J \int_{\mathbb{R}} \mathrm d t \; \sin(\varphi_{\text{cl}}(t)) \sin(\varphi_q(t)).
    \end{split}
\end{equation}
The Green's function in the Keldysh technique can be defined through the functional integral 
\begin{equation}
    i G_{ab} = \int D[\varphi_{\text{cl}}, \varphi_q] \; \varphi_a \varphi_b \;
    e^{i (S_0[\varphi] + S_{\text{int}}[\varphi])},
\end{equation}
where the indices $a$, $b$ label the components in the $\{\text{cl}, q\}$ notation and time.
Applying Wick's theorem, we obtain a correction term 
\begin{equation}
    \begin{split}
        &G_{ab} - G_{ab}^{(0)} = i G_{a c}^{(0)} \times \\
        &\times \left\langle 
            \left(
                i \frac{\delta^2 S_{\text{int}}[\varphi]}{\delta \varphi_c \delta \varphi_d} - 
                \frac{\delta S_{\text{int}}[\varphi]}{\delta \varphi_c} \frac{\delta S_{\text{int}}[\varphi]}{\delta \varphi_d}
            \right) e^{i S_\text{int}[\varphi]}
        \right\rangle_0 G_{db}^{(0)},
    \end{split}
\end{equation}
where summation over repeated indices $c, d$ is implied, and 
$\langle \cdot \rangle_0$ corresponds to averaging with the quadratic action $S_0$.
According to Dyson's equation, up to second order in the perturbation parameter $E_J$,
\begin{equation}
    \hat{G} = \hat{G}_0 + \hat{G}_0 \hat{\Sigma} \hat{G}_0 + \hat{G}_0 \hat{\Sigma} \hat{G}_0 \hat{\Sigma} \hat{G}_0 + \dots
\end{equation}
we can extract the first- and second-order self-energy parts, respectively ($\hat{\sigma}^x$ is a Pauli matrix):
\begin{equation}
    \Sigma_{1, ab} = - \left\langle 
        \frac{\delta^2 S_{\text{int}}[\varphi]}{\delta \varphi_a \delta \varphi_b}    
    \right\rangle_0 = e^{-\frac{i}{2} G^K(t = 0)} \sigma^x_{ab}, 
\end{equation}
\begin{equation}
    \begin{split}
        &\Sigma_{1, ac} G^{(0)}_{cd} \Sigma_{1, d b} + \Sigma_{2, ab} = \\
        &= -\left\langle 
            i \frac{\delta^2 S_{\text{int}}[\varphi]}{\delta \varphi_a \delta \varphi_b} S_{\text{int}}[\varphi] - 
            i \frac{\delta S_{\text{int}}[\varphi]}{\delta \varphi_a} \frac{\delta S_{\text{int}}[\varphi]}{\delta \varphi_b}
        \right\rangle_0.
    \end{split}
\end{equation}
Transforming these expressions, we obtain Eqs.~\eqref{eq:sigma_1}, \eqref{eq:sigma_2}.

For further analysis, we need the asymptotics of $S_{\pm}(t)$ (here we use massive (regularized) Green's functions):
\begin{equation}
    \begin{split}
        &S_{\pm}(t) = e^{-i (G^K_m(0) \pm G^K_m(t))} \sin(G^R_m(t)) \approx \\
        &\approx - e^{-i G^K_m(0)} 
        \begin{cases}
            e^{\pm \frac{8}{g} \frac{1}{(m t)^2}} \sin\left(\frac{4\pi}{g} e^{-mt}\right), \quad t \gg 1/m, \\
            |m t|^{\pm \frac{8}{g}} e^{\pm \frac{8 \gamma}{g}} \sin\left(\frac{4\pi}{g}\right), \quad 1/\Lambda \ll t \ll 1/m, \\
            \frac{4\pi}{g}e^{\mp i G^K_m(0)} \Lambda t, \quad \Lambda t \ll 1.
        \end{cases}
    \end{split}
    \label{eq:S_pm(t)}
\end{equation}
Here we see that $S_+(t) \equiv 0$ as $m \rightarrow 0$. 

In the case $m = 0$, the function $S_-$ allows one to calculate, through Eq.~\eqref{eq:sigma_2}, the frequency-dependent part of $\Sigma_2(\omega)$, 
which takes the form~\eqref{Sigma_2} and contains the coefficient 
\begin{equation}
    \begin{split}
        C(g) &= - \frac{i}{2}  e^{- \frac{4 i \pi}{g}} \left(\frac{\pi e^{-\gamma}}{g}\right)^{\frac{8}{g}}
        \Gamma\left(1 - \frac{8}{g}\right) \sin\left(\frac{4\pi}{g}\right).
    \end{split}
\end{equation}
In the strong-coupling case, this coefficient is (see Eqs.~\eqref{eq:duality}, \eqref{eq:Sigma2_dual})
\begin{equation}
    \begin{split}
        &C^{\text{dual}}(g) = - i  e^{- \frac{i \pi g}{2}} \frac{128 \pi}{g} \times \\
        &\times
        \left|\frac{\pi g}{16}\right|^{\frac{g}{2} - 1}
        e^{-\frac{\gamma g}{2}} \Gamma\left(1 - \frac{g}{2}\right) \sin\left(\frac{\pi g}{4}\right).
    \end{split}
\end{equation}

Next, we show that for $4 < g < 8$ the perturbation series breaks down for the dressed Green's function $G_m^{(0)}$.
We start our analysis from Eq.~\eqref{eq:sigma_2}. 
Let us transform this expression for further convenience:
\begin{equation}
    \begin{split}
        &\Sigma_2(t) = \Sigma^{(2)}_{1,1}(t) + \Sigma^{(2)}_{1,2}(t) + \Sigma^{(2)}_{2,1}(t) + 
        \Sigma^{(2)}_{2,2}(t) + \Sigma^{(2)}_{3}(t), \\ 
        &\Sigma^{(2)}_{1,1}(t) = -\frac{1}{2} \delta(t) \int_{0}^{+\infty}
        \mathrm d \tau \; e^{i (G^K_{m}(\tau) - G^K_{m}(0))} \sin(G^R_{m}(\tau)), \\
        &\Sigma^{(2)}_{1,2}(t) = \frac{1}{2} \delta(t) \int_{0}^{+\infty}
        \mathrm d \tau \; e^{-i (G^K_{m}(\tau) + G^K_{m}(0))} \sin(G^R_{m}(\tau)), \\
        &\Sigma^{(2)}_{2,1}(t) = \frac{1}{2} \sin(G^R_{m}(t))
        e^{i (G^K_{m}(t) - G^K_{m}(0))}, \\
        &\Sigma^{(2)}_{2,2}(t) = \frac{1}{2} \sin(G^R_{m}(t))
        e^{-i (G^K_{m}(t) + G^K_{m}(0))}, \\
        &\Sigma^{(2)}_3(t) = - G^R_{m}(t) e^{-i G^K_m(0)}.
    \end{split}
\end{equation}
The local term is obtained by taking $\omega \rightarrow +0 \Leftrightarrow t \rightarrow +\infty$; thus, 
we should use the large‑time approximation, i.e. $t \gg \frac{1}{m}$:
\begin{equation}
    \begin{split}
        &\Sigma^{(2)}_{1r} = \Sigma^{(2)}_{1, 1} + \Sigma^{(2)}_{1, 2} \sim \frac{1}{E_C} \left(\frac{m}{\Lambda}\right)^{\frac{8}{g} - 1}.
    \end{split}
\end{equation}
For $m \ll \omega \ll \Lambda$ we compare the powers of 
$E_J$ for the non‑local contributions:
\begin{equation}
    \begin{split}
        &E_J^2 \Sigma^{(2)}_{2, 1}(m) \sim E_J^{2 + \frac{8 - g}{g - 4}}, \\
        &E_J^2 \Sigma^{(2)}_{2, 1}(g E_C/\pi) \sim E_J^{2}, \\
        &E_J^2 \Sigma^{(2)}_{2, 2}(m) \sim E_J^{2 + \frac{8 - g}{g - 4}}, \\
        &E_J^2 \Sigma^{(2)}_{2, 2}(g E_C/\pi) \sim E_J^{2 + \frac{16}{g - 4}}, \\
        &E_J^2 \Sigma^{(2)}_{3}(m) \sim E_J^{2 + \frac{8 - g}{g - 4}}, \\
        &E_J^2 \Sigma^{(2)}_{3}(g E_C/\pi) \sim E_J^{2 + \frac{8}{g - 4}}.
    \end{split}
\end{equation}
Thus, $\Sigma^{(2)}_{2, 2}$ and $\Sigma^{(2)}_{3}$ can be neglected. 
So we have shown that to second order in $E_J$ we can retain only $\Sigma^{(2)}_{1r}$ and 
$\Sigma^{(2)}_{2,1}$, since they are the leading terms in the perturbation parameter $E_J$.

We should also compare all terms in $\Sigma_2(\omega)$ with $m$:
\begin{equation}
\begin{split}
        &m \sim E_J^{\frac{g}{g-4}}, \\
        &E_J^2 \Sigma^{(2)}_{1r} \sim E_J^{2 + \frac{8 - g}{g - 4}}, \\
        &E_J^2 \Sigma^{(2)}_{2,1}(g E_C/\pi) \sim E_J^2, \\ 
        &E_J^2 \Sigma^{(2)}_{2, 1} (m) \sim E_J^{2 + \frac{8 - g}{g - 4}}.
\end{split}
\label{eq:app_Omega_Sigma_compare}
\end{equation}
Here we find that the mass term is small for $g \in (4; 8)$ 
(in comparison with $E_J^2 \Sigma^{(2)}_{2,1}(\omega)$ 
and 
$E_J^2 \Sigma^{(2)}_{2,1}(m)$) 
in the region $m \ll \omega \ll \Lambda$.
Thus, for $g > 8$ we have mass contributions coming from $m$ and $16 E_C E_J^2\Sigma^{(2)}_{1r}$.
This analysis is correct for intermediate frequencies $m \ll \omega \ll \Lambda$.

For small frequencies $\omega \ll \Omega^2/\alpha$, we should calculate
\begin{equation}
    \begin{split}
        &\Sigma^{(2)}_{2r}(0 < \omega \ll m) = \Sigma^{(2)}_{2,1}(\omega) + \Sigma^{(2)}_{2,2}(\omega) + \Sigma^{(2)}_3(\omega) \approx \\
        &\approx 
        - \left(\frac{\pi^2 m}{g^2 E_C^2}\right)^{\frac{8}{g} - 1}
        \frac{128 \pi}{g^3} \frac{2 + e \operatorname{Ei}(-1)}{6e} - \\
        &-
        \frac{i \omega}{\left(\frac{g E_C}{\pi}\right)^2} \frac{64 \pi|\operatorname{Ei}(-1)|}{g^3} \left(\frac{\pi m}{g E_C}\right)^{\frac{8}{g} - 2}.
    \end{split}
    \label{eq:Sigma_2_low_frequency}
\end{equation}

\newpage

\bibliographystyle{unsrt}
\bibliography{references}

@article{Schmid1983,
  author  = {Schmid, A.},
  title   = {Diffusion and localization in a dissipative quantum system},
  journal = {Physical Review Letters},
  volume  = {51},
  number  = {17},
  pages   = {1506--1509},
  year    = {1983},
  doi     = {10.1103/PhysRevLett.51.1506}
}

@article{Bulgadaev1984,
  author  = {Bulgadaev, S. A.},
  title   = {Phase diagram of a dissipative quantum system with a periodic potential},
  journal = {JETP Letters},
  volume  = {39},
  pages   = {315--319},
  year    = {1984}
}

@article{SchonZaikin1990,
  author  = {Sch{\"o}n, G. and Zaikin, A. D.},
  title   = {Quantum coherent effects, phase transitions, and the dissipative dynamics of ultra small tunnel junctions},
  journal = {Physics Reports},
  volume  = {198},
  number  = {5-6},
  pages   = {237--412},
  year    = {1990},
  doi     = {10.1016/0370-1573(90)90035-S}
}

@article{Kuzmin2019NP,
  author  = {Kuzmin, R. and Mencia, R. and Grabon, N. and Mehta, N. and Lin, Y.-H. and Manucharyan, V. E.},
  title   = {Quantum electrodynamics of a superconductor--insulator phase transition},
  journal = {Nature Physics},
  volume  = {15},
  pages   = {930--934},
  year    = {2019},
  doi     = {10.1038/s41567-019-0553-1}
}

@article{Kuzmin2019NPJ,
  author  = {Kuzmin, R. and Mehta, N. and Grabon, N. and Mencia, R. and Manucharyan, V. E.},
  title   = {Superstrong coupling in circuit quantum electrodynamics},
  journal = {npj Quantum Information},
  volume  = {5},
  pages   = {20},
  year    = {2019},
  doi     = {10.1038/s41534-019-0134-2}
}

@article{Kuzmin2025,
  author  = {Kuzmin, Roman and Mehta, Nitish and Grabon, Nicholas and Mencia, Raymond A. and Burshtein, Amir and Goldstein, Moshe and Manucharyan, Vladimir},
  title   = {Observation of the Schmid--Bulgadaev dissipative quantum phase transition},
  journal = {Nature Physics},
  volume  = {21},
  pages   = {132--136},
  year    = {2025},
  doi     = {10.1038/s41567-024-02695-7}
}

@article{Kuzmin_scat,
  title = {Inelastic Scattering of a Photon by a Quantum Phase Slip},
  author = {Kuzmin, R. and Grabon, N. and Mehta, N. and Burshtein, A. and Goldstein, M. and Houzet, M. and Glazman, L. I. and Manucharyan, V. E.},
  journal = {Phys. Rev. Lett.},
  volume = {126},
  issue = {19},
  pages = {197701},
  numpages = {6},
  year = {2021},
  month = {May},
  publisher = {American Physical Society},
  doi = {10.1103/PhysRevLett.126.197701},
  url = {https://link.aps.org/doi/10.1103/PhysRevLett.126.197701}
}

@article{Subero2026,
  author        = {Subero, Diego and Chang, Yu-Cheng and Monteiro, Miguel and Chen, Ze-Yan and Pekola, Jukka P.},
  title         = {Revisiting dissipation-driven phase transition in a Josephson junction},
  journal       = {Physical Review Letters},
  volume        = {136},
  number        = {11},
  year          = {2026},
  doi           = {10.1103/4f3c-jv43},
  eprint        = {2509.20480},
  archivePrefix = {arXiv},
  primaryClass  = {cond-mat.mes-hall}
}

@article{Penttila2001,
  author  = {Penttil{\"a}, J. S. and Hakonen, P. J. and Sonin, E. B. and Paalanen, M. A.},
  title   = {Experiments on Dissipative Dynamics of Single Josephson Junctions},
  journal = {Journal of Low Temperature Physics},
  volume  = {125},
  number  = {3--4},
  pages   = {89--114},
  year    = {2001},
  doi     = {10.1023/A:1012971500694}
}

@article{HakonenSonin2021,
  author    = {Hakonen, Pertti J. and Sonin, Edouard B.},
  title     = {Comment on ``Absence of a Dissipative Quantum Phase Transition in Josephson Junctions''},
  journal   = {Physical Review X},
  volume    = {11},
  number    = {1},
  pages     = {018001},
  year      = {2021},
  doi       = {10.1103/PhysRevX.11.018001}
}

@article{Burshtein2024,
  title = {Photon-instanton scattering in a superconducting circuit: Beyond the very high impedance regime},
  author = {Burshtein, A. and Oganesov, V. and Goldstein, M. and Glazman, L. I.},
  journal = {Physical Review B},
  year = {2024},
  doi = {10.1103/36gc-zdyh}
}

@article{Burshtein2020,
  title = {Photon-Instanton Collider Implemented by a Superconducting Circuit},
  author = {Burshtein, A. and Kuzmin, R. and Manucharyan, V. E. and Goldstein, M.},
  journal = {Physical Review Letters},
  year = {2020},
  doi = {10.1103/PhysRevLett.126.137701}
}

@article{Houzet2020,
  title = {Critical Fluorescence of a Transmon at the Schmid Transition},
  author = {Houzet, M. and Glazman, L. I.},
  journal = {Physical Review Letters},
  year = {2020},
  doi = {10.1103/physrevlett.125.267701}
}

@article{Paris_diaganalization,
  title = {Resilience of the quantum critical line in the Schmid transition},
  author = {Paris, Nicolas and Giacomelli, Luca and Daviet, Romain and Ciuti, Cristiano and Dupuis, Nicolas and Mora, Christophe},
  journal = {Phys. Rev. B},
  volume = {111},
  issue = {6},
  pages = {064509},
  numpages = {11},
  year = {2025},
  month = {Feb},
  publisher = {American Physical Society},
  doi = {10.1103/PhysRevB.111.064509},
  url = {https://link.aps.org/doi/10.1103/PhysRevB.111.064509}
}

@article{Pechenezhskiy2020,
  author  = {Pechenezhskiy, Ivan V. and Mencia, Raymond A. and Nguyen, Long B. and Lin, Yen-Hsiang and Manucharyan, Vladimir E.},
  title   = {The superconducting quasicharge qubit},
  journal = {Nature},
  year    = {2020},
  volume  = {585},
  number  = {7825},
  pages   = {368--371},
  doi     = {10.1038/s41586-020-2687-9},
  url     = {https://doi.org/10.1038/s41586-020-2687-9},
  issn    = {1476-4687}
}

@book{Kamenev2023,
  author    = {Kamenev, Alex},
  title     = {Field Theory of Non-Equilibrium Systems},
  edition   = {2},
  publisher = {Cambridge University Press},
  year      = {2023},
  doi       = {10.1017/9781108769266},
  isbn      = {9781108769266},
  url       = {https://doi.org/10.1017/9781108769266}
}

@article{arutyunov,
  title={Superconductivity in one dimension},
  author={Arutyunov, K Yu and Golubev, Dmitri S and Zaikin, Andrej Dmitievi{\v{c}}},
  journal={Physics Reports},
  volume={464},
  number={1-2},
  pages={1--70},
  year={2008},
  publisher={Elsevier}
}

@article{mooij,
  title={Propagating plasma mode in thin superconducting filaments},
  author={Mooij, JE and Sch{\"o}n, Gerd},
  journal={Physical review letters},
  volume={55},
  number={1},
  pages={114},
  year={1985},
  publisher={APS}
}

@article{RSZ17,
  title={Quantum phase fluctuations and density of states in superconducting nanowires},
  author={Radkevich, Alexey and Semenov, Andrew G and Zaikin, Andrei D},
  journal={Physical Review B},
  volume={96},
  number={8},
  pages={085435},
  year={2017},
  publisher={APS}
}

@article{RSZ19,
  title={Quantum fluctuations and phase coherence in superconducting nanowires},
  author={Radkevich, Alexey and Semenov, Andrew G and Zaikin, Andrei D},
  journal={Physical Review B},
  volume={100},
  number={1},
  pages={014520},
  year={2019},
  publisher={APS}
}

@article{ALRSZ,
  title={Superconducting insulators and localization of Cooper pairs},
  author={Arutyunov, Konstantin Yu and Lehtinen, Janne S and Radkevich, Alexey and Semenov, Andrew G and Zaikin, Andrei D},
  journal={Communications Physics},
  volume={4},
  number={1},
  pages={146},
  year={2021},
  publisher={Nature Publishing Group UK London}
}

@article{RSZJSNM20,
  title={Topology-controlled phase coherence and quantum fluctuations in superconducting nanowires},
  author={Radkevich, Alexey and Semenov, Andrew G and Zaikin, Andrei D},
  journal={Journal of Superconductivity and Novel Magnetism},
  volume={33},
  number={8},
  pages={2335--2339},
  year={2020},
  publisher={Springer}
}

@article{kampf1987quantum,
  title={Quantum effects and the dissipation by quasiparticle tunneling in arrays of Josephson junctions},
  author={Kampf, Arno and Sch{\"o}n, Gerd},
  journal={Physical Review B},
  volume={36},
  number={7},
  pages={3651},
  year={1987},
  publisher={APS}
}

@article{bulgadaev1984phase,
  title={Phase diagram of a dissipative quantum system},
  author={Bulgadaev, SA},
  journal={ZhETF Pisma Redaktsiiu},
  volume={39},
  pages={264},
  year={1984}
}

@article{guinea1985diffusion,
  title={Diffusion and localization of a particle in a periodic potential coupled to a dissipative environment},
  author={Guinea, F and Hakim, V and Muramatsu, A},
  journal={Physical review letters},
  volume={54},
  number={4},
  pages={263},
  year={1985},
  publisher={APS}
}

\end{document}